\documentclass[twocolumn,showpacs,preprintnumbers]{revtex4}
\usepackage{amsmath}
\usepackage{amsfonts}
\usepackage{graphicx}
\usepackage{dcolumn}
\usepackage{bm}

\begin{document}

\title{Multi-photon Scattering Theory and Generalized Master Equations}
\author{Tao Shi$^{1}$, Darrick E. Chang$^{2}$, and J. Ignacio Cirac$^{1}$}
\affiliation{$^{1}$Max-Planck-Institut f\"{u}r Quantenoptik, Hans-Kopfermann-Strasse. 1,
85748 Garching, Germany \\
$^{2}$ICFO-Institut de Ciencies Fotoniques, Mediterranean Technology Park,
08860 Castelldefels (Barcelona), Spain}
\date{\today }

\begin{abstract}
We develop a scattering theory to investigate the multi-photon transmission
in a one-dimensional waveguide in the presence of quantum emitters. It is
based on a path integral formalism, uses displacement transformations, and
does not require the Markov approximation. We obtain the full time-evolution
of the global system, including the emitters and the photonic field. Our
theory allows us to compute the transition amplitude between arbitrary
initial and final states, as well as the S-matrix of the asymptotic in- and
out- states. For the case of few incident photons in the waveguide, we also
re-derive a generalized master equation in the Markov limit. We compare the
predictions of the developed scattering theory and that with the Markov
approximation. We illustrate our methods with five examples of few-photon
scattering: (i) by a two-level emitter, (ii) in the Jaynes-Cummings model;
(iii) by an array of two-level emitters; (iv) by a two-level emitter in the
half-end waveguide; (v) by an array of atoms coupled to Rydberg levels. In
the first two, we show the application of the scattering theory in the
photon scattering by a single emitter, and examine the correctness of our
theory with the well-known results. In the third example, we analyze the
condition of the Markov approximation for the photon scattering in the array
of emitters. In the forth one, we show how a quantum emitter can generate
entanglement of out-going photons. Finally, we highlight the interplay
between the phenomenon of electromagnetic-induced transparency and the
Rydberg interaction, and show how this results in a rich variety of
possibilities in the quantum statistics of the scattering photons.
\end{abstract}

\pacs{03.65.Nk, 42.50.-p, 11.55.-m, 72.10.Fk}
\author{}
\maketitle


\section{Introduction}

The exploration of quantum optical systems characterized by strong
photon-photon interactions has inspired a lot of research and extensive
studies recently \cite{Imamoglu,Kimble,SPDarrick,qno}. Those systems provide
us with a versatile platform to investigate the generation and transport of
non-classical light, as well as the behavior of single-photon sources \cite%
{edsp,sptwolevel,semisp} and switches \cite%
{SPDarrick,switch1,switch2,switch3,SPcosk}. Those are the basic ingredients
in quantum-optical \cite{QO} and quantum information devices \cite{QI,QNet}.

The manipulation of non-classical light typically requires devices
displaying either strong nonlinearities \cite{qno}, or quantum interference
effects \cite{inter1,inter2}. Among many other phenomena, they give rise to
peculiar quantum statistical behavior of the emitted or scattered photons,
like anti-bunching \cite{QObook} in the generation of single photons or
photon pairs \cite{pair1,pair2}. Such devices are being investigated in
different incarnations, including cavity QED \cite%
{Imamoglu,Kimble,inter1,inter2,Fan,LSZ,JCShi,inter3,inter4}, solid state
\cite{edsp,semisp,solid,diamond,NV,nanowire}, and circuit QED systems \cite%
{SPcircuit,SPcQED}. At the many-particle level, the strong interaction
between the nonlinear devices and the photons results in the generation of
many-body states, which can be studied in terms of dissipative versions of
quantum spin models \cite{spinmodel}. In particular, the investigations of
the atomic steady state and the photon transmission properties reveal a rich
variety of quantum phases \cite{MP1,MP2} and photon statistics.

In order to characterize how atomic (or any other) non-linear devices can be
used to create and manipulate photonic states, one can analyze the
transmission spectra and the photon statistics, for instance, in terms of
the second order correlation function of the photons emitted under the
presence of weak driving light. This analysis is typically addressed through
an input-output theory relating the correlation functions of the emitted
photons to those of the atomic system in steady state \cite{QObook}. Those
can be determined using a master equation approach, based on the Born-Markov
approximation. This approach has proven to be very successful in most of the
experimentally relevant situations. Even though that is a very good
approximation for most models in quantum optics, in the presence of several
emitters, or in certain regimes its use may not be justified.

In order to analyze the transmission properties exactly, several elegant
approaches \cite%
{Fan,LSZ,io,Fanfi,RydDarrick,WPQC,PBTF,MBBS,tEIT,NLCT,NPWP,1DQS} have been
developed for the few photon scattering process, where the Born-Markov
approximation is not involved. The exact analysis of single and two photon
transmissions was first addressed through the Bethe ansatz approach \cite%
{Fan}, which is equivalent to the Lippmann-Schwinger scattering theory. This
approach establishes the exact scattering matrix ($S$-matrix) between the
in- and out- asymptotic states of photons, which determines the transmission
spectrum and the second order correlation function of outgoing photons. It
turns out that the Bethe ansatz approach is very successful in the
two-photon scattering by a single emitter with simple structure, e.g., the
two-level emitter, however, the generalization to the photon transmission by
several emitters is difficult. The approach \cite{io} based on the
input-output theory is able to provide the exact $S$-matrix for the two
photon scattering by two emitters. Here, since a closed set of motion
equations for emitter operators are required, the exact $S$-matrix can only
be obtained in some very limited cases, e.g., two emitters with simple
structures.

The systematic approach to the exact $S$-matrix in the multi-photon
scattering process is based on quantum field theory. Through either the
Lehmann-Symanzik-Zimmerman (LSZ) reduction formalism \cite{LSZ} or the
input-output theory, the exact $S$-matrix is related to the time ordered
correlation function of emitter operators, which can be obtained by the
functional integration of the photonic bath modes. The functional
integration provides an efficient approach to analyze the photon statistical
properties in the multi-photon scattering by emitters with complicated
structures, e.g., the Jaynes-Cummings (JC) system \cite{JCShi} and the
atom-coupled whispering gallery resonator \cite{inter4}. However, the exact $%
S$-matrix is only able to describe the evolution of photonic bath in the
asymptotic limit, thus, it fails to depict the transient dynamics of
emitters, e.g., the single-photon detection in the superconducting edge
sensor \cite{edge1,edge2,edge3}. Recently, a generalized input-output theory
\cite{RydDarrick} was established by the subtle combination of the
input-output theory and the quantum regression theorem, which allowed us to
investigate both the transient dynamics and the photon statistics in the
asymptotic limit. Here, a generalized master equation is obtained to
describe the transient dynamics of several emitters under the presence of
few incident photons. The generalized input-output theory and master
equation involve the Markov approximation, thus, they are not able to
provide the exact results. The use of the Markov approximation in the
presence of several emitters need to be justified.

In this paper, based on the path integral formalism, we develop the
scattering theory to characterize the full time evolution of the global
system (including the emitters and the photonic bath) exactly for the
few-photon scattering by several emitters with complicated structures. Our
theory provides the transition amplitudes from the arbitrary initial state
to the corresponding final state without the Markov approximation, where
both the quantum statistics of scattering photons and the transient dynamics
of emitters can be analyzed exactly. In the Markovian limit, the exact
results from our theory perfectly agree with those from the quantum
regression approach, which justifies the validity of the Markov
approximation. Similarly, the generalized master equation is re-produced to
describe the dynamics of emitters in the presence of few incident photons.
Here, we emphasize that the generalized master equation establishes the
close relation between the properties of photons emitted under the present
of weak driving light and those in the few-photon scattering by the emitters.

Using the two paradigmatic examples, i.e., the few-photon transmission to
the two-level emitter and the JC system, we examine the correctness of our
theory in the photon scattering by the single emitter. Here, the developed
scattering theory re-produces the well-known results \cite%
{Fan,LSZ,JCShi,Fanec} for the transmission spectra and the second order
correlation function. For the few-photon scattering by several emitters, we
investigate the condition of the Markov approximation in detail. In
particular, we show that the Markov approximation is valid under certain
conditions relating the bandwidth of the dynamics and the distance of
separation between emitters. In the non-Markovian regime, the exact results
exhibit peculiar features associated with retardation of pulses between
emitters.

By the developed scattering theory, we explore some novel phenomena in two
new situations. For the photon scattering by a single two-level emitter in
the half-end waveguide, we show how the two-level emitter can generate
entanglement of out-going photons. For the photon transmission in the array
of atoms under conditions of electromagnetically induced transparency (EIT)
\cite{rEIT} and coupled to Rydberg levels, we highlight the EIT phenomenon
and the Rydberg interaction. Here, the second order correlation functions of
emitted photons, the co-propagation of dark polaritons and the collision of
the counter-propagating polaritons in the transient process are analyzed,
which show a rich variety of the quantum statistics of photons and Rydberg
excitations. The developed scattering theory is proven to be a very
efficient and systematical approach to investigate the quantum statistics of
photons in the array of the interacting emitters with complicated structures.

The paper is organized as follows. In Sec. II, the five models are
introduced, which are the two-level emitter and the JC system coupled to the
photonic waveguide, the two-level emitter in the half-end waveguide, and the
photon scattering by an array of two-level atoms and EIT atoms coupled to
Rydberg-levels. In Sec. III, the developed scattering theory is established,
where the exact $S$-matrix relates the asymptotic incident and final states.
For the transient process, the generalized master equation is derived to
characterize the time evolution of the quantum emitters. In Sec. IV, two
paradigmatic examples, i.e., the photon transmission to the two-level
emitter and the JC system, are used to examine our theory and illustrate how
our method works in the photon scattering by the single emitter. In Sec. V,
we analyze the validity of the Markov approximation for the photon
scattering by several emitters. In Sec. VI, we show that the generation of
entanglement between two scattering photons by a single two-level emitter in
the half-end waveguide. In Sec. VII, the photon transmission to an array of
EIT atoms coupled to Rdyberg levels is investigated. In Sec. VIII, the
results are summarized with the outlook.

\section{Scattering models}

In this section, we introduce the models we are going to use in order to
investigate the transmission of waveguide photons interacting with quantum
systems. Those could be single emitters (e.g., a multi-level atom \cite%
{Fan,SPDarrick,SPcosk,LSZ,3level,f2level}, an atom coupled to a cavity mode
\cite{Imamoglu,Kimble}), or an array of emitters coupled, for instance, to
Rydberg levels \cite{RydDarrick,JCarray}.

The model Hamiltonian has the form $H=H_{\mathrm{w}}+H_{\mathrm{sys}}+H_{%
\mathrm{\Gamma }}+H_{\mathrm{\Gamma }_{f}}$ and contains four parts: (a) The
free propagation of photons in the waveguide is described by the Hamiltonian%
\begin{equation}
H_{\mathrm{w}}=\sum_{k}[(k_{0}+k)r_{k}^{\dagger
}r_{k}+(k_{0}-k)l_{k}^{\dagger }l_{k}],
\end{equation}%
where $r_{k}$ ($r_{k}^{\dagger }$) and $l_{k}$ ($l_{k}^{\dagger }$) are the
annihilation (creation) operators for the right- and left- moving modes with
the central frequency $k_{0}$ in the waveguide; (b) The system Hamiltonian $%
H_{\mathrm{sys}}$ describes the quantum emitters, and will be explicitly
given latter for different examples; (c) The interaction between the
waveguide photons and the emitters is given by the Hamiltonian%
\begin{equation}
H_{\mathrm{\Gamma }}=\sum_{i}[\sqrt{\Gamma _{r,i}}r^{\dagger }(x_{i})+\sqrt{%
\Gamma _{l,i}}l^{\dagger }(x_{i})]O_{i}+\mathrm{H.c.},  \label{Hgx}
\end{equation}%
where the left- and right-moving photon fields $r(x_{i})=%
\sum_{k}r_{k}e^{i(k+k_{0})x_{i}}/\sqrt{L}$ and $l(x_{i})=%
\sum_{k}l_{k}e^{i(k-k_{0})x_{i}}/\sqrt{L}$ couple to the operator $O_{i}$ of
the $i$-th emitter at the position $x_{i}$ with coupling strengths $\Gamma
_{r(l),i}^{1/2}$. In momentum space, the interaction Hamiltonian reads%
\begin{equation}
H_{\mathrm{\Gamma }}=\sum_{k}(r_{k}^{\dagger }O_{k,+}+l_{k}^{\dagger
}O_{k,-})+\mathrm{H.c.},
\end{equation}%
where the collective operators $O_{k,\pm }=\sum_{i}\sqrt{\Gamma _{i}}%
O_{i}e^{-i(k\pm k_{0})x_{i}}/\sqrt{L}$, and we focus on the symmetric
coupling case, i.e., $\Gamma _{r,i}=\Gamma _{l,i}\equiv \Gamma _{i}$.
Without loss of generality, we choose the position of the first emitter $i=1$
at the origin $x_{1}=0$; (d) Apart from the decay to the waveguide, there
may exist other decay channels to free space (spontaneous emission outside
the waveguide). We describe this through the Hamiltonian $H_{\mathrm{\Gamma }%
_{f}}$. Here, each emitter couples to some system operator $\tilde{O}_{i}$
with a JC-type coupling (\ref{Hgx}).

\begin{figure}[tbp]
\includegraphics[bb=18 243 578 753, width=8 cm, clip]{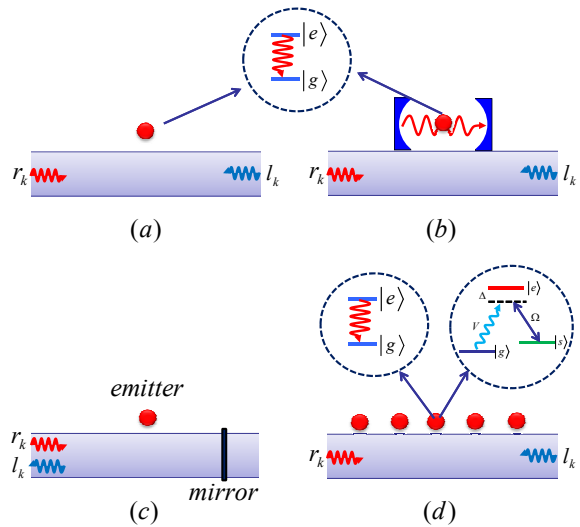}
\caption{(Color Online) Four examples of the problem studied here: (a)
Single two-level system coupled to a waveguide; (b) JC system coupled to a
waveguide; (c) Single two-level system coupled to a one-sided waveguide; (d)
An array of atoms coupled to a waveguide.}
\label{fig1}
\end{figure}

We now present some relevant examples for the system Hamiltonian $H_{\mathrm{%
sys}}$, and which will be analyzed in detail in the following sections. As
shown in Fig. \ref{fig1}a, the simplest example consists of a two-level
system with the energy level spacing $\omega _{e}=k_{0}$ between states $%
\left\vert e\right\rangle $ and $\left\vert g\right\rangle $. The
Hamiltonian is $H_{\mathrm{sys}}=\omega _{e}\left\vert e\right\rangle
\left\langle e\right\vert $ and $O_{i=1}=\tilde{O}_{i}=O=\tilde{O}%
=\left\vert g\right\rangle \left\langle e\right\vert $.

The second example is the JC system shown in Fig. \ref{fig1}b. The
Hamiltonian%
\begin{equation}
H_{\mathrm{sys}}=\omega _{c}a^{\dagger }a+\omega _{e}\left\vert
e\right\rangle \left\langle e\right\vert +g(a^{\dagger }\left\vert
g\right\rangle \left\langle e\right\vert +\mathrm{H.c.})
\end{equation}%
describes a single two-level system, with the energy level spacing $\omega
_{e}$, coupled to a cavity mode of frequency $\omega _{c}$. The annihilation
operator $a$ corresponds to the cavity mode, and the coupling constant is $g$%
. Here, the cavity mode directly couples to the waveguide, i.e., $%
O_{i=1}=O=a $. The excitations of cavity field and the two-level system can
decay into free space; thus, $\tilde{O}_{i}$ would be $a$ and $\left\vert
g\right\rangle \left\langle e\right\vert $.

We analyze the validity of the Markov approximation and the retardation
effect using the third example. Here, as shown in Fig. \ref{fig1}d, the
waveguide photons couple to an array of two-level emitters with frequency $%
\omega _{e}=k_{0}$ and lattice spacing $d$. The emitter Hamiltonian is%
\begin{equation}
H_{\mathrm{sys}}=\sum_{i}(\omega _{e}b_{i}^{\dagger }b_{i}+\frac{1}{2}%
U_{0}b_{i}^{\dagger }b_{i}^{\dagger }b_{i}b_{i}),
\end{equation}%
where the hardcore boson $b_{i}$ is introduced to describe a two-level
emitter, and the hardcore behavior is obtained in the limit $%
U_{0}\rightarrow \infty $. The waveguide photons couple to the collective
modes $O_{k,\pm }=\sqrt{\Gamma /L}\sum_{i}e^{-i(k\pm k_{0})x_{i}}b_{i}$,
where the decay rates $\Gamma _{r(l),i}$ are taken to be the constant $%
\Gamma _{r(l),i}=\Gamma $. The excitation decays into the free space, i.e., $%
\tilde{O}_{i}=b_{i}$.

The fourth example is a single emitter in front of a mirror, as shown in
Fig. \ref{fig1}c. The system Hamiltonian is $H_{\mathrm{sys}}=H_{\mathrm{%
emitter}}+\omega _{b}b^{\dagger }b$, and the collective operators are%
\begin{eqnarray}
O_{k,+} &=&\sqrt{\frac{\Gamma }{L}}Oe^{-i(k+k_{0})x_{0}}+\sqrt{\frac{\Gamma
_{b}}{L}}b,  \notag \\
O_{k,-} &=&\sqrt{\frac{\Gamma }{L}}Oe^{-i(k-k_{0})x_{0}}+\sqrt{\frac{\Gamma
_{b}}{L}}b,  \label{O}
\end{eqnarray}%
where the boson mode $b$ is introduced to describe the mirror in the limit $%
\Gamma _{b}\rightarrow \infty $. The decay of the emitter excitation into
free space is characterized by $\tilde{O}_{i}=\left\vert g\right\rangle
\left\langle e\right\vert $.

The fifth example is an array of EIT atoms coupled to Rydberg levels, with
lattice spacing $d$, as it is schematically shown in Fig. \ref{fig1}d. The
emitter Hamiltonian is%
\begin{eqnarray}
H_{\mathrm{sys}} &=&\sum_{i}[\omega _{e}e_{i}^{\dagger }e_{i}+\omega
_{s}s_{i}^{\dagger }s_{i}+(\Omega e^{-i\omega _{d}t}e_{i}^{\dagger }s_{i}+%
\mathrm{H.c.})]  \notag \\
&&+H_{\mathrm{HC}}+\frac{1}{2}\sum_{ij}U_{ij}s_{i}^{\dagger }s_{j}^{\dagger
}s_{j}s_{i}.  \label{HR}
\end{eqnarray}%
Here, we use hardcore bosons $e$ and $s$ to describe a Rydberg-EIT atom.
There are three atomic levels: the ground state $\left\vert g\right\rangle $%
, the excited state $\left\vert e\right\rangle $ with frequency $\omega _{e}$%
, and the Rydberg state $\left\vert s\right\rangle $ with frequency $\omega
_{s}$. The hardcore behavior is obtained through%
\begin{equation}
H_{\mathrm{HC}}=\frac{U_{0}}{2}\sum_{i}(e_{i}^{\dagger }e_{i}+s_{i}^{\dagger
}s_{i})(e_{i}^{\dagger }e_{i}+s_{i}^{\dagger }s_{i}-1)
\end{equation}%
in the limit $U_{0}\rightarrow \infty $, which projects out the states $%
\left\vert ee\right\rangle $, $\left\vert es\right\rangle $, and $\left\vert
ss\right\rangle $ of each atom with double occupations. The atoms in the
Rydberg state $\left\vert s\right\rangle $ experience long-range
interactions, $U_{ij}$, \cite{RydH}. For instance, we can take the
Van-der-Waals interaction $U_{ij}=C_{6}/\left\vert x_{i}-x_{j}\right\vert
^{6}$, the dipolar interaction $U_{ij}=C_{3}/\left\vert
x_{i}-x_{j}\right\vert ^{3}$, or the uniform interaction $U_{ij}=C_{0}$ \cite%
{RydDarrick}. The transition between states $\left\vert e\right\rangle $ and
$\left\vert s\right\rangle $ is induced by a classical field of frequency $%
\omega _{d}$ and corresponding Rabi frequency $\Omega $. The waveguide
photons couple to the collective modes $O_{k,\pm }=\sqrt{\Gamma }%
\sum_{i}e^{-i(k\pm k_{0})x_{i}}e_{i}/\sqrt{L}$, where the decay rates $%
\Gamma _{r(l),i}$ are taken to be the constant $\Gamma _{r(l),i}=\Gamma $.
The $e$-excitations can decay into the free space, i.e., $\tilde{O}%
_{i}=e_{i} $.

For convenience, we transform the Hamiltonian in the rotating frame $%
r_{k}\rightarrow r_{k}e^{-ik_{0}t}$ and $l_{k}\rightarrow l_{k}e^{-ik_{0}t}$%
. The system operators are also transformed correspondingly such that the
total Hamiltonian is time-independent. As a result, in the rotating frame, $%
H_{\mathrm{w}}=\sum_{k}k(r_{k}^{\dagger }r_{k}-l_{k}^{\dagger }l_{k})$, and
the frequencies in the system Hamiltonian are replaced by the detunings.
This will be used in the five examples presented in the following sections.

\section{General formalism}

In this section, we show the general formalism to study the full dynamics of
the global system during the few photon scattering process. In Secs. IIIA
and IIIB, we briefly review the results based on the Markov approximation in
Ref. \cite{RydDarrick}. Here, a different method, i.e., the displacement
transformation, is introduced to relate the transition amplitude between
arbitrary initial and finial states with the correlators of emitter
operators. Based on the quantum regression theorem, these correlators are
obtained by the effective Hamiltonian of the emitters, where the quantum
regression theorem requires the Markov approximation. In order to obtain the
exact result and examine the validity of the Markov approximation, in Sec.
IIIC we use the path integral formalism to derive the exact transition
amplitude, where the non-Markovian effects are taken into account.

In both approaches, the $S$-matrix of the asymptotic in- and out- states and
the transition amplitudes in the transient process are obtained, which agree
with each other in the Markov limit. It turns out that for the single
emitter coupled to the photon with linear dispersion, the Markovian result
is proven to be exact by the path integral approach. For the photon
scattering by several emitters, the validity of the Markov approximation is
analyzed in Sec. V in detail. In the Markov limit, both approaches give rise
to the generalized master equation \cite{RydDarrick} governing the dynamics
of the emitters during the scattering process, where the generalized master
equation establishes a close relation between the transient behaviors of
emitters in the presence of few incident photons and the dynamics of
emitters under the weak driving light.

\subsection{$S$-matrix by quantum regression theorem}

In this section, based on the quantum regression theorem, the displacement
transformation is used to derive the transition amplitude between arbitrary
initial and final states, where the quantum regression theorem involves the
Markov approximation.

We introduce the transition amplitude%
\begin{equation}
\mathcal{A}(T)=\text{ }_{\mathrm{sys}}\left\langle \varphi _{\mathrm{out}%
}\right\vert \left\langle \mathrm{out}\right\vert
e^{-iH(t_{f}-t_{i})}\left\vert \mathrm{in}\right\rangle \left\vert \varphi _{%
\mathrm{in}}\right\rangle _{\mathrm{sys}}
\end{equation}%
from the initial state $\left\vert \mathrm{in}\right\rangle \left\vert
\varphi _{\mathrm{in}}\right\rangle _{\mathrm{sys}}=\left\vert
0\right\rangle _{\mathrm{free}}\left\vert \psi _{\mathrm{in}}\right\rangle _{%
\mathrm{w}}\left\vert \varphi _{\mathrm{in}}\right\rangle _{\mathrm{sys}}$
to the final state $\left\vert \mathrm{out}\right\rangle \left\vert \varphi
_{\mathrm{out}}\right\rangle _{\mathrm{sys}}=\left\vert 0\right\rangle _{%
\mathrm{free}}\left\vert \psi _{\mathrm{out}}\right\rangle _{\mathrm{w}%
}\left\vert \varphi _{\mathrm{out}}\right\rangle _{\mathrm{sys}}$ during the
time $T=t_{f}-t_{i}$. At the instant $t_{i}$, the waveguide photons are in
the state $\left\vert \psi _{\mathrm{in}}\right\rangle _{\mathrm{w}}$, and
the initial state of the emitters is $\left\vert \varphi _{\mathrm{in}%
}\right\rangle _{\mathrm{sys}}=\gamma _{\mathrm{in}}^{\dagger }\left\vert
G\right\rangle _{\mathrm{sys}}$, where $\left\vert G\right\rangle _{\mathrm{%
sys}}$ and $\gamma _{\mathrm{in}}^{\dagger }$ denote the ground state and
some creation operator of the emitters, respectively. Similarly, at the
instant $t_{f}$, the corresponding final states are $\left\vert \varphi _{%
\mathrm{out}}\right\rangle _{\mathrm{sys}}=\gamma _{\mathrm{out}}^{\dagger
}\left\vert G\right\rangle _{\mathrm{sys}}$ and $\left\vert \psi _{\mathrm{%
out}}\right\rangle _{\mathrm{w}}$, respectively. Here, we focus on the
transition process without excitations leaking to the free space, and the
initial and final states of the free space are the vacuum state $\left\vert
0\right\rangle _{\mathrm{free}}$.

The initial and final states of waveguide photons can be generally written as%
\begin{eqnarray}
\left\vert \psi _{\mathrm{in}}\right\rangle _{\mathrm{w}}
&=&\sum_{\{n_{k\alpha }\}}\psi _{\mathrm{in}}(\{n_{k\alpha
}\})\prod_{k}\left\vert n_{k\alpha }\right\rangle ,  \notag \\
\left\vert \psi _{\mathrm{out}}\right\rangle _{\mathrm{w}}
&=&\sum_{\{m_{k\alpha }\}}\psi _{\mathrm{out}}(\{m_{k\alpha
}\})\prod_{k}\left\vert m_{k\alpha }\right\rangle ,
\end{eqnarray}%
where $\{n_{k\alpha }\}=\{n_{k_{1}\alpha },n_{k_{2}\alpha },...\}$ and $%
\{m_{k\alpha }\}=\{m_{k_{1}\alpha },m_{k_{2}\alpha },...\}$ are the number
distribution of photons with different momenta $k_{i}$ in the initial and
final states, respectively, and $\alpha =r,l$ denote the right- and
left-moving modes. We notice that the relation%
\begin{equation}
\left\vert n_{k\alpha }\right\rangle =\lim_{J_{k\alpha }\rightarrow 0}\frac{1%
}{\sqrt{n_{k\alpha }!}}\frac{\partial ^{n_{k\alpha }}}{\partial J_{k\alpha
}^{n_{k\alpha }}}\left\vert J_{k\alpha }\right\rangle
\end{equation}%
between the Fock state and the unnormalized coherent state $\left\vert
J_{k\alpha }\right\rangle =\sum_{n_{k}}J_{k\alpha }^{n_{k}}\left\vert
n_{k\alpha }\right\rangle /\sqrt{n_{k\alpha }!}$ leads to the coherent
representation of the initial and final states%
\begin{eqnarray}
\left\vert \psi _{\mathrm{in}}\right\rangle _{\mathrm{w}} &=&\mathcal{F}_{%
\mathrm{in}}\left\vert \{J_{k\alpha }\}\right\rangle ,  \notag \\
\left\vert \psi _{\mathrm{out}}\right\rangle _{\mathrm{w}} &=&\mathcal{F}_{%
\mathrm{out}}\left\vert \{J_{k\alpha }\}\right\rangle ,  \label{fio}
\end{eqnarray}%
where%
\begin{eqnarray}
\mathcal{F}_{\mathrm{in}} &=&\lim_{\{J_{k\alpha }\}\rightarrow
0}\sum_{\{n_{k\alpha }\}}\psi _{\mathrm{in}}(\{n_{k\alpha }\})\prod_{k\alpha
}\frac{1}{\sqrt{n_{k\alpha }!}}\frac{\delta ^{n_{k\alpha }}}{\delta
J_{k\alpha }^{n_{k\alpha }}}, \\
\mathcal{F}_{\mathrm{out}} &=&\lim_{\{J_{k\alpha }\}\rightarrow
0}\sum_{\{m_{k\alpha }\}}\psi _{\mathrm{out}}(\{m_{k\alpha
}\})\prod_{k\alpha }\frac{1}{\sqrt{m_{k\alpha }!}}\frac{\delta ^{m_{k\alpha
}}}{\delta J_{k\alpha }^{m_{k\alpha }}}.  \notag
\end{eqnarray}

In terms of Eq. (\ref{fio}), the transition amplitude reads%
\begin{equation}
\mathcal{A}(T)=\mathcal{F}_{\mathrm{out}}^{\ast }\mathcal{F}_{\mathrm{in}}%
\mathcal{A}_{J}(T),  \label{At}
\end{equation}%
where the transition amplitude%
\begin{equation}
\mathcal{A}_{J}(T)=\text{ }_{\mathrm{sys}}\left\langle \varphi _{\mathrm{out}%
}\right\vert _{\mathrm{b}}\left\langle \{J_{k\alpha }\}\right\vert
e^{-iHT}\left\vert \{J_{k\alpha }\}\right\rangle _{\mathrm{b}}\left\vert
\varphi _{\mathrm{in}}\right\rangle _{\mathrm{sys}},  \label{TA}
\end{equation}%
and $\left\vert \{J_{k\alpha }\}\right\rangle _{\mathrm{b}}=\left\vert
0\right\rangle _{\mathrm{free}}\left\vert \{J_{k\alpha }\}\right\rangle $.
The transition amplitude (\ref{TA}) can be evaluated by either the quantum
regression theorem or the path integral approach.

We first show the result from the quantum regression theorem, where the
Markov approximation is required. In the interacting picture, the time
evolution operator $\mathcal{U}=\mathcal{T}\exp
[-i\int_{t_{i}}^{t_{f}}dtH(t)]$ is determined by the Hamiltonian $H(t)=H_{%
\mathrm{sys}}+H_{\mathrm{\Gamma }_{f}}+H_{\mathrm{\Gamma }}(t)$, where $%
\mathcal{T}$ is the time-ordering operator and the interaction part $H_{%
\mathrm{\Gamma }}(t)=e^{iH_{\mathrm{w}}t}H_{\mathrm{\Gamma }}e^{-iH_{\mathrm{%
w}}t}$ is%
\begin{equation}
H_{\mathrm{\Gamma }}(t)=\sum_{k}(r_{k}^{\dagger
}O_{k,+}e^{ikt}+l_{k}^{\dagger }O_{k,-}e^{-ikt})+\mathrm{H.c.}.
\end{equation}%
In terms of $\mathcal{U}$, the amplitude (\ref{TA}) reads%
\begin{equation}
\mathcal{A}_{J}(T)=\text{ }_{\mathrm{sys}}\left\langle \varphi _{\mathrm{out}%
}\right\vert _{\mathrm{b}}\left\langle \{J_{k\alpha ,\mathrm{out}%
}\}\right\vert \mathcal{U}\left\vert \{J_{k\alpha ,\mathrm{in}%
}\}\right\rangle _{\mathrm{b}}\left\vert \varphi _{\mathrm{in}}\right\rangle
_{\mathrm{sys}},
\end{equation}%
where $J_{k\alpha ,\mathrm{in}}=J_{k\alpha }e^{i\varepsilon _{k\alpha
}t_{i}} $, $J_{k\alpha ,\mathrm{out}}=J_{k\alpha }e^{i\varepsilon _{k\alpha
}t_{f}}$, the dispersion relations $\varepsilon _{k,\alpha }=\sigma _{\alpha
}k$, and $\sigma _{r(l)}=\pm $ for the right- and left- moving modes,
respectively.

The displacement transformation $U$: $\left\vert \{J_{k\alpha
}\}\right\rangle =e^{\sum_{k\alpha }\left\vert J_{k\alpha }\right\vert
^{2}/2}U\left\vert \{0_{k\alpha }\}\right\rangle $ is introduced to rewrite
the amplitude%
\begin{eqnarray}
\mathcal{A}_{J}(T) &=&\exp (\sum_{k,\alpha =r,l}\left\vert J_{k,\alpha
}\right\vert ^{2}e^{-i\varepsilon _{k,\alpha }T})\times \\
&&\text{ }_{\mathrm{sys}}\left\langle \varphi _{\mathrm{out}}\right\vert _{%
\mathrm{b}}\left\langle \{0_{k\alpha }\}\right\vert \mathrm{\tilde{U}}%
\left\vert \{0_{k\alpha }\}\right\rangle _{\mathrm{b}}\left\vert \varphi _{%
\mathrm{in}}\right\rangle _{\mathrm{sys}}  \notag
\end{eqnarray}%
where the state of the waveguide photon is transformed to the vacuum state,
and the time evolution operator $\mathrm{\tilde{U}}=\mathcal{T}%
e^{-i\int_{t_{i}}^{t_{f}}dt[H(t)+H_{\mathrm{d}}(t)]}$ is determined by $H(t)$
and the driving term%
\begin{eqnarray}
H_{\mathrm{d}}(t) &=&\sum_{k,\alpha =r,l}[J_{k,\alpha }^{\ast }O_{k,\sigma
_{\alpha }}e^{-i\sigma _{\alpha }k(t_{f}-t)}  \label{Hd} \\
&&+J_{k,\alpha }O_{k,\sigma _{\alpha }}^{\dagger }e^{-i\sigma _{\alpha
}k(t-t_{i})}].  \notag
\end{eqnarray}

It follows from Eq. (\ref{At}) that the functional derivative of $\mathcal{A}%
_{J}(T)$ determines the transition amplitude $\mathcal{A}(T)$, which is
composed of Fourier transforms of time ordered correlators%
\begin{equation}
\mathcal{G}(T)=\left\langle \mathcal{T}\gamma _{\mathrm{out}%
}(t_{f})O_{1}(t_{1})...O_{n}(t_{n})\gamma _{\mathrm{in}}^{\dagger
}(t_{i})\right\rangle
\end{equation}%
on the ground state $\left\vert \{0_{k}\}\right\rangle _{\mathrm{b}%
}\left\vert G\right\rangle _{\mathrm{sys}}$. Here, $O_{j}(t_{j})=\mathcal{U}%
^{\dagger }(t)O_{j}\mathcal{U}(t)$ is given by the emitter operators $%
O_{j}=O_{k_{j},\pm }$ and $O_{k_{j},\pm }^{\dagger }$.

By the quantum regression theorem, as shown in Ref. \cite{RydDarrick}, the
bath degree of freedom can be traced out and the correlator%
\begin{equation*}
\mathcal{G}(T)=\text{ }_{\mathrm{sys}}\left\langle G\right\vert \mathcal{T}%
\gamma _{\mathrm{out}}(t_{f})O_{\mathrm{eff},1}(t_{1})...O_{\mathrm{eff}%
,n}(t_{n})\gamma _{\mathrm{in}}^{\dagger }(t_{i})\left\vert G\right\rangle _{%
\mathrm{sys}}
\end{equation*}%
becomes the average value of emitter operators $O_{\mathrm{eff},j}(t)=%
\mathcal{U}_{\mathrm{eff}}^{\dagger }(t)O_{j}\mathcal{U}_{\mathrm{eff}}(t)$
on the ground state $\left\vert G\right\rangle _{\mathrm{sys}}$. The
time-evolution operator $\mathcal{U}_{\mathrm{eff}}(t)=\exp (-iH_{\mathrm{eff%
}}t)$ is given by the non-Hermitian effective Hamiltonian $H_{\mathrm{eff}%
}=H_{\mathrm{sys}}+H_{\mathrm{decay}}$, where%
\begin{equation}
H_{\mathrm{decay}}=-i\sum_{i}\Gamma _{f,i}\tilde{O}_{i}^{\dagger }\tilde{O}%
_{i}-i\sum_{ij}\sqrt{\Gamma _{i}\Gamma _{j}}O_{i}^{\dagger
}O_{j}e^{ik_{0}\left\vert x_{i}-x_{j}\right\vert },  \label{Heff1}
\end{equation}%
and $\Gamma _{f,i}$ is decay rate to the free space of the emitter at the
position $x_{i}$.

Based on the quantum regression theorem, each term in $\mathcal{A}(T)$ is
related to the correlator of emitter operators $O_{\mathrm{eff},j}(t)$
governed by the effective Hamiltonian $H_{\mathrm{eff}}$. The re-summation
of these terms results in $\mathcal{A}(T)=\mathcal{F}_{\mathrm{out}}^{\ast }%
\mathcal{F}_{\mathrm{in}}\mathcal{A}_{J}(T)$ with the compact form%
\begin{equation}
\mathcal{A}_{J}(T)=\text{ }_{\mathrm{sys}}\left\langle \varphi _{\mathrm{out}%
}\right\vert \mathcal{U}_{\mathrm{eff}}\left\vert \varphi _{\mathrm{in}%
}\right\rangle _{\mathrm{sys}}e^{\sum_{k,\alpha =r,l}\left\vert J_{k,\alpha
}\right\vert ^{2}e^{-i\varepsilon _{k,\alpha }T}},  \label{TA2}
\end{equation}%
where $\mathcal{U}_{\mathrm{eff}}=\mathcal{T}e^{-i\int_{t_{i}}^{t_{f}}dt(H_{%
\mathrm{eff}}+H_{\mathrm{d}})}$. The equations (\ref{At}) and (\ref{TA2})
establish the relation between the transition amplitude $\mathcal{A}(T)$ and
the time ordered correlators of emitters governed by the effective
Hamiltonian $H_{\mathrm{eff}}$.

By different choices of $\gamma _{\mathrm{in}}$, $\gamma _{\mathrm{out}}$, $%
t_{i}$, $t_{f}$, $\mathcal{F}_{\mathrm{in}}$ and $\mathcal{F}_{\mathrm{out}}$%
, the dynamics of the global system can be fully characterized. Hereafter,
we refer the choice of $\gamma _{\mathrm{in(out)}}$, $t_{(i,f)}$, and $%
\mathcal{F}_{\mathrm{in(out)}}$ as the boundary condition. For instance, $%
\mathcal{A}(T)$ with the boundary condition $\gamma _{\mathrm{in}}=\gamma _{%
\mathrm{out}}=I$ (identity operator), $t_{i}\rightarrow -\infty $, and $%
t_{f}\rightarrow \infty $ leads to the photonic $S$-matrix. For the boundary
condition $\gamma _{\mathrm{in}}=I$, $\gamma _{\mathrm{out}}\neq I$, $%
t_{i}=0 $, and $t_{f}=T$, $\mathcal{A}(T)$ describes how the incident
photons transform to the emitter excitations and propagate in the transient
regime. For the boundary condition $\gamma _{\mathrm{in}}\neq I$, $\gamma _{%
\mathrm{out}}=I$, $t_{i}=0$, and $t_{f}=T$, the behaviors of spontaneous and
stimulated emissions can be investigated by $\mathcal{A}(T)$. Using Eqs. (%
\ref{At}), (\ref{Heff1}), and (\ref{TA2}), together with different boundary
conditions, we shall study the few photon scattering process in the five
models in Secs. IV-VII.

\subsection{Generalized master equation}

In order to study the dynamics of emitters during the scattering, one can
either use the transition amplitude (\ref{At}) with proper boundary
conditions or derive the master equation of emitter reduced density operator
by tracing out the photonic degree of freedom. In quantum optics, the
initial states of the photonic bath are usually the vacuum state, the
thermal state, and Gaussian states, where the conventional master equation
is obtained based on the Born-Markov approximation.

During the scattering process, the initial state of photons is dramatically
changed. For instance, the single incident photon resonant with the
transition energy of the two-level emitter is totally reflected, as
predicted by the single photon scattering theory \cite%
{Fan,SPDarrick,SPcosk,LSZ}, where the photon in the initial asymptotic state
$r_{k}^{\dagger }\left\vert 0\right\rangle $ is totally scattered to the
photon in the out-going asymptotic state $l_{-k}^{\dagger }\left\vert
0\right\rangle $. However, the conventional master equation assumes that the
initial state of the photonic bath is unchanged, and the state of the global
system is always the product state of the emitter and the photonic bath \cite%
{QObook}. This assumption contradicts with the exact result of the single
photon scattering theory, thus, the conventional master equation breaks
down. The generalized master equation is required to investigate the
dynamics of emitters in the few-photon scattering process.

For the initial state of few photons in the waveguide, the evolution of the
system state is described by the reduced density matrix $\rho _{s}(T)=Tr_{%
\mathrm{bath}}[\mathcal{U}\rho (0)\mathcal{U}^{\dagger }]$, where in terms
of the coherent state $\left\vert \{J_{k\alpha }\}\right\rangle $%
\begin{equation}
\rho (0)=\mathcal{F}_{\mathrm{in}}^{\ast }\mathcal{F}_{\mathrm{in}}[\rho _{%
\mathrm{sys}}(0)\otimes \left\vert \{J_{k\alpha }\}\right\rangle _{\mathrm{b}%
}\left\langle \{J_{k\alpha }\}\right\vert ].
\end{equation}%
The displacement transformation $U$ relates the density matrix%
\begin{equation}
\rho _{s}(T)=\mathcal{F}_{\mathrm{in}}^{\ast }\mathcal{F}_{\mathrm{in}%
}e^{\sum_{k\alpha }\left\vert J_{k\alpha }\right\vert ^{2}}\rho _{J}(T)
\label{r}
\end{equation}%
to the generating density matrix%
\begin{equation}
\rho _{J}(T)=Tr_{\mathrm{bath}}[\mathcal{U}_{J}\rho _{\mathrm{sys}%
}(0)\otimes \left\vert \{0_{k\alpha }\}\right\rangle \left\langle
\{0_{k\alpha }\}\right\vert \mathcal{U}_{J}^{\dagger }],
\end{equation}%
where $\mathcal{U}_{J}=U^{\dagger }\mathcal{U}U$ describes the evolution of
emitters under the driving field. In $\rho _{J}(T)$, the initial state of
photons is transformed to the vacuum state, thus, the conventional master
equation can be used to describe the time evolution of $\rho _{J}(T)$. Under
the Markov approximation, the master equation governing the evolution of $%
\rho _{J}(T)$ reads%
\begin{equation}
\partial _{T}\rho _{J}(T)=-i[\tilde{H}_{\mathrm{sys}},\rho _{J}(T)]+\mathcal{%
L}\rho _{J}(T),  \label{gr}
\end{equation}%
where the initial condition is $\rho _{J}(0)=\rho _{\mathrm{sys}}(0)$. The
displacement transformation induces the driving term in%
\begin{eqnarray}
&&\tilde{H}_{\mathrm{sys}}=H_{\mathrm{sys}}+\sum_{ij}\sqrt{\Gamma _{i}\Gamma
_{j}}O_{i}^{\dagger }O_{j}\sin (k_{0}\left\vert x_{i}-x_{j}\right\vert )
\notag \\
&&+\sum_{k,\alpha }(J_{k,\alpha }^{\ast }O_{k,\sigma _{\alpha }}e^{i\sigma
_{a}kT}+J_{k,\alpha }O_{k,\sigma _{\alpha }}^{\dagger }e^{-i\sigma _{a}kT}),
\label{hd}
\end{eqnarray}%
and the Lindblad operator is%
\begin{eqnarray}
&&\mathcal{L}\rho _{J}(T)=2\sum_{ij}\sqrt{\Gamma _{i}\Gamma _{j}}O_{i}\rho
_{J}(T)O_{j}^{\dagger }\cos [k_{0}(x_{i}-x_{j})]  \notag \\
&&-\sum_{ij}\sqrt{\Gamma _{i}\Gamma _{j}}\cos
[k_{0}(x_{i}-x_{j})]\{O_{i}^{\dagger }O_{j},\rho _{J}(T)\}  \label{ld} \\
&&+\sum_{i}\Gamma _{f,i}[2\tilde{O}_{i}\rho _{J}(T)\tilde{O}_{i}^{\dagger
}-\{\tilde{O}_{i}^{\dagger }\tilde{O}_{i},\rho _{J}(T)\}].  \notag
\end{eqnarray}

The generalized master Eqs. (\ref{r}) and (\ref{gr}) are the main results in
this section, which lead to some significant results: (a) In principle, once
$\rho _{J}(T)$ is obtained, $\rho _{s}(T)$ is determined through Eq. (\ref{r}%
), which describes the transient dynamics of emitters during the scattering
processes. In practice, the evolution of emitters for the few incident
photons can be studied by the perturbative expansion of the classical
sources $J_{k,\alpha }$ and $J_{k,\alpha }^{\ast }$. For instance, for the
single incident photon, Eq. (\ref{r}) contains at most the second order
derivative $\delta ^{2}\rho _{J}(T)/\delta J_{p\alpha ^{\prime }}^{\ast
}\delta J_{k\alpha }$, the second order perturbative expansion of $%
J_{k,\alpha }$ and $J_{k,\alpha }^{\ast }$ in $\rho _{J}(T)$ determines $%
\rho _{s}(T)$. Here, the zero order and second order contributions lead to
the emitter reduced density matrix for the single incident photon. The zero
order contribution $\rho _{\mathrm{sys}}(0)$ describes the unchanged emitter
state without interacting with the photon, while the second order
contribution $\mathcal{F}_{\mathrm{in}}^{\ast }\mathcal{F}_{\mathrm{in}}\rho
_{J}(T)$ describes the dynamics of emitters responding to the single
incident photon. Similarly, for two incident photons, the reduced density
matrix can be determined by the forth order expansion of $J_{k,\alpha }$ and
$J_{k,\alpha }^{\ast }$.

(b) In quantum optics, we are interested on the transmission properties of
weak probe light to emitters, i.e., the transmission spectrum and the
quantum statistics. We notice that Eq. (\ref{gr}) just describes the
emitters under the classical probe light with the strength $J_{k\alpha }$.
If the driving field is weak, we can solve Eq. (\ref{gr}) by the
perturbative expansion of $J_{k,\alpha }$ and $J_{k,\alpha }^{\ast }$. As we
discussed above, the second and forth order perturbative expansions describe
the transient dynamics of emitters in the single and two photon scattering
processes. As a result, the quantum properties of out-going photons and
emitters under the weak driving light can be studied by\ the few photon
scattering theory, where only the time evolution in the forward path is
involved, as shown in Eq. (\ref{TA2}). We shall show in Sec. VII that for
the emitters with complicated structures the scattering theory is still able
to provide the analytic results, which perfectly agree with the numerical
results from the master equation approach. The few photon scattering theory
enables us to analytically investigate the quantum properties of photons
from emitters under the weak diving field.

(c) If we set the external source to be zero, Eq. (\ref{gr}) agrees with the
master equation \cite{spinmodel} for the photonic bath initially in the
ground state.

\subsection{Exact $S$-matrix by path integral approach}

The equations (\ref{At}), (\ref{Heff1}), and (\ref{TA2}) in Sec. IIIA relate
the transition amplitude to the correlators of emitters governing by the
effective Hamiltonian, where under the Markov approximation the quantum
regression theorem leads to the instantaneous effective Hamiltonian. In this
section, we show the exact result of the transition amplitude (\ref{At}),
from which the result obtained by the quantum regression theorem is proven
to exact for the single emitter coupled to the waveguide. The validity of
the Markov approximation and some non-Markov effects for several emitters
coupled to the waveguide will be investigated by the exact transition
amplitude in Sec. VI.

Following the procedure of path integral formalism, we discretize the
evolution time $T$ by $N\rightarrow \infty $ instants $\{t_{1}....t_{N}\}$
and insert coherent state basis at each instant $t_{i}$. By integrating out
the free space modes we obtain%
\begin{eqnarray}
\mathcal{A}_{J}(T) &=&\int D[\mathrm{system}]\gamma _{\mathrm{out}%
}(t_{f})\gamma _{\mathrm{in}}^{\ast }(t_{i})e^{iS_{\mathrm{sys}}}  \notag \\
&&\times \int D[\alpha _{k},\alpha _{k}^{\ast }]e^{\sum_{k,\alpha
}J_{k,\alpha }^{\ast }\alpha _{k}(t_{f})}e^{iS},
\end{eqnarray}%
where the action%
\begin{equation}
S=\int_{t_{i}}^{t_{f}}dt\{\sum_{k}[r_{k}^{\ast }(i\partial
_{t}-k)r_{k}+l_{k}^{\ast }(i\partial _{t}+k)l_{k}]-H_{\mathrm{\Gamma }}\}
\label{AS}
\end{equation}%
describes free propagation of the right- and left- moving photons in the
waveguide and the interaction with the emitters, and $\int D[\mathrm{system}%
] $ is the integral over the emitter field. The action of the emitter is
\begin{equation}
S_{\mathrm{sys}}=S_{\mathrm{sys}}^{(0)}+i\int_{t_{i}}^{t_{f}}dt\sum_{i}%
\Gamma _{f,i}\tilde{O}_{i}^{\dagger }\tilde{O}_{i},  \label{Ss}
\end{equation}%
where the first term is the action of emitters, and the second term
describes the decay to the free space.

By $\delta S/\delta r_{k}^{\ast }=0$ and $\delta S/\delta l_{k}^{\ast }=0$,
the classical motion equations read%
\begin{eqnarray}
(i\partial _{t}-k)r_{k,\mathrm{cl}}-O_{k,+} &=&0,  \notag \\
(i\partial _{t}+k)l_{k,\mathrm{cl}}-O_{k,-} &=&0,
\end{eqnarray}
which give the classical paths%
\begin{eqnarray}
r_{k,\mathrm{cl}}(t)
&=&J_{k,r}e^{-ik(t-t_{i})}-i\int_{t_{i}}^{t}dsO_{k,+}(s)e^{-ik(t-s)},  \notag
\\
l_{k,\mathrm{cl}}(t)
&=&J_{k,l}e^{ik(t-t_{i})}-i\int_{t_{i}}^{t}dsO_{k,-}(s)e^{ik(t-s)}.
\end{eqnarray}

Following the saddle point method, we expand the photon fields $r_{k}=r_{k,%
\mathrm{cl}}+\delta r_{k}$ and $l_{k}=l_{k,\mathrm{cl}}+\delta l_{k}$ by the
quantum fluctuation fields $\delta r_{k}$ and $\delta l_{k}$ around the
classical paths, and integrate out the fluctuation fields in $\mathcal{A}%
_{J}(T)$. Due to the quadratic form of the action (\ref{AS}),%
\begin{eqnarray}
\mathcal{A}_{J}(T) &=&\exp (\sum_{k,\alpha }\left\vert J_{k,\alpha
}\right\vert ^{2}e^{-i\sigma _{\alpha }kT})  \label{AJt} \\
&&\times \int D[\mathrm{system}]\gamma _{\mathrm{out}}(t_{f})\gamma _{%
\mathrm{in}}^{\ast }(t_{i})e^{iS_{\mathrm{eff}}+iS_{J}}  \notag
\end{eqnarray}%
is obtained exactly, where the nonlocal effective action $S_{\mathrm{eff}%
}=S_{\mathrm{sys}}+S_{\mathrm{re}}$ of emitters is determined by%
\begin{eqnarray}
S_{\mathrm{re}}
&=&i\int_{t_{i}}^{t_{f}}dt\int_{t_{i}}^{t}ds\sum_{k}[O_{k,+}^{\ast
}(t)O_{k,+}(s)e^{-ik(t-s)}  \notag \\
&&+O_{k,-}^{\ast }(t)O_{k,-}(s)e^{ik(t-s)}],  \label{re}
\end{eqnarray}%
and the source terms $S_{J}=-\int_{t_{i}}^{t_{f}}dtH_{\mathrm{d}}(t)$.

The equations (\ref{AJt}) and (\ref{re}) are the central results in this
section. It follows from Eq. (\ref{At}) that the functional derivatives of $%
\mathcal{A}_{J}(T)$ lead to the transition amplitude $\mathcal{A}(T)$. We
also notice that, apart from the phase factor, $\mathcal{A}_{J}(T)$ is the
generating functional \cite{QFT} of the emitter correlators, namely, the
functional derivatives of $\mathcal{A}_{J}(T)$ give the correlation
functions of emitter operators. As a result, the transition amplitude $%
\mathcal{A}(T)$ is determined by the correlation functions of emitter
fields, where the effective action $S_{\mathrm{eff}}$ is generally time
nonlocal.

For the single emitter coupled to the waveguide with the linear dispersion,
the action $S_{\mathrm{eff}}$ turns out to be time-local, and the effective
Hamiltonian $H_{\mathrm{eff}}$ is deduced. Here, the renormalized part (\ref%
{re}) has the time-local form%
\begin{equation}
S_{\mathrm{re}}=i\Gamma \int_{t_{i}}^{t_{f}}dtO^{\ast }(t)O(t),
\end{equation}%
and effective Hamiltonian%
\begin{equation}
H_{\mathrm{eff}}=H_{\mathrm{sys}}-i\Gamma _{f}\tilde{O}^{\dagger }\tilde{O}%
-i\Gamma O^{\dagger }O  \label{Hes}
\end{equation}%
of emitters is obtained. In fact, the functional integral in Eq. (\ref{AJt})
is just the amplitude $_{\mathrm{sys}}\left\langle \varphi _{\mathrm{out}%
}\right\vert \mathcal{U}_{\mathrm{eff}}\left\vert \varphi _{\mathrm{in}%
}\right\rangle _{\mathrm{sys}}$ in Eq. (\ref{TA2}), thus, the result based
on the quantum regression theorem agrees with the exact result (\ref{AJt}).

For several emitters, the renormalization part%
\begin{equation*}
S_{\mathrm{re}}=i\int_{t_{i}}^{t_{f}}dt\sum_{ij}\sqrt{\Gamma _{i}\Gamma _{j}}%
O_{i}^{\ast }(t)O_{j}(t-\left\vert x_{i}-x_{j}\right\vert
)e^{ik_{0}\left\vert x_{i}-x_{j}\right\vert }.
\end{equation*}%
is time nonlocal. If the wavelength of frequency fluctuations around $k_{0}$
is much larger than the size $Nd$ of the emitter array, the time delay
effect can be neglected, which leads to the time-local action%
\begin{equation}
S_{\mathrm{re}}=i\int_{t_{i}}^{t_{f}}dt\sum_{ij}\sqrt{\Gamma _{i}\Gamma _{j}}%
O_{i}^{\ast }(t)O_{j}(t)e^{ik_{0}\left\vert x_{i}-x_{j}\right\vert },
\end{equation}%
and the effective Hamiltonian (\ref{Heff1}). In this limit, the result (\ref%
{TA2}) agrees with the exact result (\ref{AJt}) for several emitters.

In the end of this section, we briefly summarize the results. We use two
approaches to obtain the transition amplitude and the generalized master
equation for the photon scattering process, where the derivation of the
generalized master equation (\ref{gr}) from the path integral approach is
left in the Appendix. The relation between the few photon scattering by
emitters and the emitters under weak driving light is established. For the
single emitter case, the result (\ref{TA2}) based on the quantum regression
theorem is proven to be exact.\ For several emitters, in the Markovian
limit, the result (\ref{TA2}) agrees with the exact result from the path
integral approach.

\section{Paradigmatic examples}

In this section, as paradigmatic examples, the few-photon scattering by the
single emitter is studied by the scattering theory developed in Sec. III.
The emitter is chosen to be the two-level emitter or the JC system, where
the decay to the free space is neglected, i.e., $\Gamma _{f}=0$. For the
two-level emitter, we investigate the transmission spectra and the quantum
statistics of out-going photons for the single and two incident photons. For
the emitter initially in the excited state without the incident photon, the
spontaneous emission of the emitter is studied. The transient process, i.e.,
the response of the two-level emitter in the ground state to the single
photon wave-packet, is analyzed by the exact transition amplitude of the
emitter in the excited state. For the stimulated emission of the two-level
emitter initially in the excited state, the wave-packet shape of the single
incident photon is designed, such that the probability of stimulated
emission is maximal. For the JC system, we show the transmission spectra and
the quantum statistics of scattering photons. By these two examples, we
examine our theory by the comparison with the well-known results in Refs.
\cite{Fan,LSZ,JCShi,Fanec}, which justifies the correctness of the developed
scattering theory.

\subsection{Two-level emitter}

For the two-level emitter, $O=\sigma ^{-}$ and the effective Hamiltonian $H_{%
\mathrm{eff}}=H_{\mathrm{sys}}-i\Gamma \sigma ^{+}\sigma ^{-}$ follows from
Eq. (\ref{Hes}), where $H_{\mathrm{sys}}$ vanishes in the rotating frame
since the transition frequency $\omega _{e}=k_{0}$ is resonant with the
central frequency of the waveguide. It follows from Eq. (\ref{Hd}) that the
external source term is%
\begin{eqnarray}
S_{J} &=&-\sqrt{\frac{\Gamma }{L}}\int_{t_{i}}^{t_{f}}dt\sum_{k,\alpha
=r,l}[J_{k,\alpha }^{\ast }\sigma ^{-}(t)e^{-i\sigma _{\alpha }k(t_{f}-t)}
\notag \\
&&+J_{k,\alpha }\sigma ^{+}(t)e^{-i\sigma _{\alpha }k(t-t_{i})}].
\end{eqnarray}

For the scattering of single right-moving photon with the momentum $k$ by
the emitter in the ground state, the boundary condition in the asymptotic
limit $t_{i}\rightarrow -\infty $ is $\gamma _{\mathrm{in}}=I$ and $\mathcal{%
F}_{\mathrm{in}}=\lim_{\{J_{k}\}\rightarrow 0}\delta /\delta J_{k,r}$. By
the boundary condition $\gamma _{\mathrm{out}}=I$ and $\mathcal{F}_{\mathrm{%
out}}=\lim_{\{J_{k}\}\rightarrow 0}\delta /\delta J_{p,(r,l)}$ in the
asymptotic limit $t_{f}\rightarrow \infty $, Eqs. (\ref{At}) and (\ref{AJt})
lead to the reflection and transmission coefficients%
\begin{equation}
R_{k}\delta _{p,-k}=\frac{-i\Gamma }{k_{0}+i\Gamma }\delta _{p,-k},
\end{equation}%
and%
\begin{equation}
T_{k}\delta _{p,k}=\frac{k_{0}}{k_{0}+i\Gamma }\delta _{p,k},
\end{equation}%
in the asymptotic limits $t_{i}\rightarrow -\infty $ and $t_{f}\rightarrow
\infty $. For the scattering of two right-moving photons with momenta $k_{1}$
and $k_{2}$ by the emitter in the ground state, the boundary condition in
the asymptotic limit $t_{i}\rightarrow -\infty $ is $\gamma _{\mathrm{in}}=I$
and $\mathcal{F}_{\mathrm{in}}=\lim_{\{J_{k}\}\rightarrow 0}\delta
^{2}/\delta J_{k_{1},r}\delta J_{k_{2},r}$. By the boundary condition $%
\gamma _{\mathrm{out}}=I$ and $\mathcal{F}_{\mathrm{out}}=\lim_{\{J_{k}\}%
\rightarrow 0}\delta ^{2}/\delta J_{-p_{1},l}\delta J_{-p_{2},l}$ in the
asymptotic limit $t_{f}\rightarrow \infty $, it follows from Eqs. (\ref{At})
and (\ref{AJt}) that the $S$-matrix%
\begin{eqnarray}
S_{p_{1}p_{2};k_{1}k_{2}} &=&\Gamma ^{2}\int_{-\infty }^{+\infty }\frac{%
dt_{1}^{\prime }dt_{2}^{\prime }dt_{1}dt_{2}}{(2\pi )^{2}}%
e^{ip_{1}t_{1}^{\prime }+ip_{2}t_{2}^{\prime }-ik_{1}t_{1}-ik_{2}t_{2}}
\notag \\
&&\left\langle \mathcal{T}\sigma ^{-}(t_{1}^{\prime })\sigma
^{-}(t_{2}^{\prime })\sigma ^{+}(t_{1})\sigma ^{+}(t_{2})\right\rangle ,
\label{S}
\end{eqnarray}%
of two reflected photons with momenta $-p_{1}$ and $-p_{2}$ is given by the
time ordered correlator of spin operators. Here, the time evolution $\sigma
^{\pm }(t)=\sigma ^{\pm }e^{\pm \Gamma t}$ governed by $H_{\mathrm{eff}}$
determines the four-point correlator, which leads to the $S$-matrix%
\begin{eqnarray}
&&S_{p_{1}p_{2};k_{1}k_{2}}=R_{k_{1}}R_{k_{2}}(\delta _{p_{1}k_{1}}\delta
_{p_{2}k_{2}}+\delta _{p_{1}k_{2}}\delta _{p_{2}k_{1}})  \notag \\
&&+\frac{i\Gamma ^{2}}{\pi }\frac{\delta
_{p_{1}+p_{2},k_{1}+k_{2}}(k_{1}+k_{2}+2i\Gamma )}{(p_{1}+i\Gamma
)(p_{2}+i\Gamma )(k_{1}+i\Gamma )(k_{2}+i\Gamma )}.  \label{S2}
\end{eqnarray}

The wavefunction of two reflected photons is obtained directly from the
Fourier transform of Eq. (\ref{S2}) \cite{LSZ} as%
\begin{eqnarray}
\psi (x_{1},x_{2}) &=&\frac{1}{2}\int
dp_{1}dp_{2}e^{ip_{1}x_{1}+ip_{2}x_{2}}S_{p_{1}p_{2};k_{1}k_{2}}  \notag \\
&=&e^{iEx_{c}}R_{k_{1}}R_{k_{2}}[\cos (kx)-e^{(i\frac{1}{2}E-\Gamma
)\left\vert x\right\vert }],  \label{wf}
\end{eqnarray}%
where the total momentum $E=k_{1}+k_{2}$, the relative momentum $%
k=(k_{1}-k_{2})/2$, the center of mass coordinate $x_{c}=(x_{1}+x_{2})/2$,
and the relative momentum $x=x_{1}-x_{2}$. The results (\ref{S2}) and (\ref%
{wf}) agree with those in Ref. \cite{Fan,LSZ}.

For two photons with the resonant frequency $k_{1}=k_{2}=0$, the reflection
coefficients are $R_{k_{1}}=R_{k_{2}}=1$ and the wavefunction%
\begin{equation}
\psi (x_{c},x)=1-e^{-\Gamma \left\vert x\right\vert }.
\end{equation}%
In Sec. III, we prove that the quantum statistics of out-going photons from
the emitter under the weak driving light can be characterized by the two
photon wavefunction $\psi (x_{c},x)$. More precisely, for the weak driving
light with the zero detuning $\Delta _{d}=\omega _{d}-\omega _{e}$, the
second order correlation function $g^{(2)}(x)=\left\vert \psi
(x_{c},x)\right\vert ^{2}$ of the emitted photons can be obtained by the two
photon wavefunction $\psi (x_{c},x)$ from the scattering theory. Here, $%
g^{(2)}(x)=(1-e^{-\Gamma \left\vert x\right\vert })^{2}$ display the
anti-bunching behavior.

\begin{figure}[tbp]
\includegraphics[bb=22 471 566 727, width=8 cm, clip]{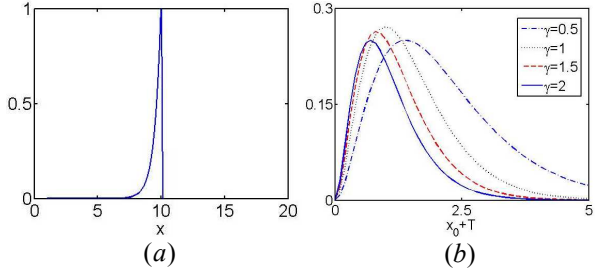}
\caption{(Color Online) Emission and absorption of a single photon by a
single emitter: (a) The spatial amplitude $|\protect\psi(x)|^{2}$ of a
right-propagating single photon wavepacket, produced by an emitter initially
prepared in the excited state at $T=0$. The wavepacket shown here is after
an evolution time $T=10/\Gamma$. (b) The probability $|A(T)|^2$ of the
emitter in the excited state for the incident wavepacket with the width $1/%
\protect\gamma$ and localized at $x_0$. Here, the emission rate $\Gamma=1$
into the waveguide is taken to be the unit.}
\label{fig2}
\end{figure}

The spontaneous decay of the two-level emitter in the excited state can be
investigated by the boundary conditions $\gamma _{\mathrm{in}}=\sigma ^{-}$,
$\mathcal{F}_{\mathrm{in}}=1$ and $\gamma _{\mathrm{out}}=I$, $\mathcal{F}_{%
\mathrm{out}}=\lim_{\{J_{k}\}\rightarrow 0}\delta /\delta J_{p,(r,l)}$ at
the initial and the final instants $t_{i}=0$ and $t_{f}=T$. The equations (%
\ref{At}) and (\ref{AJt}) give the amplitude%
\begin{equation}
\mathcal{A}(T)=\sqrt{\frac{\Gamma }{2\pi }}\frac{e^{-i\sigma _{\alpha
}pT}-e^{-\Gamma T}}{\sigma _{\alpha }p+i\Gamma }  \label{Ap}
\end{equation}%
to detect single out-going photon with the momentum $p$. The Fourier
transform gives the wavefunction%
\begin{eqnarray}
\psi (x) &=&\int \frac{dp}{\sqrt{2\pi }}e^{ipx}\mathcal{A}(T)  \notag \\
&=&-i\sqrt{\Gamma }e^{-\Gamma (T-\sigma _{a}x)}\theta (T-\sigma _{a}x)\theta
(\sigma _{a}x)
\end{eqnarray}%
in the coordinate space for the right- and left- moving photon. Since the
emitter couples to the left and right moving modes symmetrically, the
wavepackets of the right and left moving photons are symmetric with respect
to the origin. As shown in Fig. \ref{fig2}a, $\left\vert \psi (x)\right\vert
^{2}$ for the right moving photon exhibits the Lorentzian shape with the
different widths $1/\Gamma $.

The reverse process of the spontaneous decay is the response of the
two-level emitter in the ground state to the single photon wavepacket $f_{%
\mathrm{in}}(k)$. Here, the boundary conditions are $\gamma _{\mathrm{in}}=I$%
, $\mathcal{F}_{\mathrm{in}}=\lim_{\{J_{k}\}\rightarrow 0}\int dkf_{\mathrm{%
in}}(k)\delta /\delta J_{k,r}$ and $\gamma _{\mathrm{out}}=\sigma ^{-}$ and $%
\mathcal{F}_{\mathrm{out}}=1$ at the instants $t_{i}=0$ and $t_{f}=T$,
respectively. The equations (\ref{At}) and (\ref{AJt}) result in the
amplitude%
\begin{eqnarray}
\mathcal{A}(T) &=&-i\sqrt{\frac{\Gamma }{2\pi }}\int dkf_{\mathrm{in}%
}(k)\int_{0}^{T}dte^{-ikt}\left\langle \sigma ^{-}(T)\sigma
^{+}(t)\right\rangle  \notag \\
&=&\sqrt{\frac{\Gamma }{2\pi }}\int dkf_{\mathrm{in}}(k)\frac{%
e^{-ikT}-e^{-\Gamma T}}{k+i\Gamma },
\end{eqnarray}%
of the emitter in the excited state. For the right-moving photon wavepacket%
\begin{equation}
f_{\mathrm{in}}(k)=\sqrt{\frac{\gamma }{\pi }}\frac{e^{-ikx_{0}}}{k+i\gamma }
\end{equation}%
with the width $1/\gamma $ initially at the position $x_{0}<0$, the residue
theorem gives%
\begin{equation}
\mathcal{A}(T)=\frac{\sqrt{2\gamma \Gamma }}{\Gamma -\gamma }[e^{-\Gamma
(x_{0}+T)}-e^{-\gamma (x_{0}+T)}]\theta (x_{0}+T).  \label{AT}
\end{equation}%
In Fig. \ref{fig2}b, we show $\left\vert \mathcal{A}(T)\right\vert ^{2}$ of
the emitter in the excited state for different widths $\gamma $.

In the stimulated emission, the two-level emitter is initially prepared in
the excited state at the instant $t_{i}=0$, and the wavepacket of single
right-moving photon is designed to realize the maximal probability of
emitting two right-moving photons in the asymptotic limit $t_{f}\rightarrow
\infty $.

With the initial and final boundary conditions $\gamma _{\mathrm{in}}=\sigma
^{-}$, $\mathcal{F}_{\mathrm{in}}=\lim_{\{J_{k}\}\rightarrow 0}\int dkf_{%
\mathrm{in}}(k)\delta /\delta J_{k,r}$ and $\gamma _{\mathrm{out}}=I$ $%
\mathcal{F}_{\mathrm{out}}=\lim_{\{J_{k}\}\rightarrow 0}\delta ^{2}/\delta
J_{p_{1},r}\delta J_{p_{2},r}$, Eqs. (\ref{At}) and (\ref{AJt}) result in
the $S$-matrix%
\begin{eqnarray}
S_{p_{1}p_{2},k} &=&\sqrt{\frac{\Gamma }{2\pi }}\int dkf_{\mathrm{in}%
}(k)[\delta _{kp_{1}}\frac{1}{p_{2}+i\Gamma }  \notag \\
&&+\frac{\Gamma }{2\pi }\frac{1}{p_{1}+i\Gamma }\frac{1}{p_{1}-k+i0^{+}}%
\frac{1}{p_{1}+p_{2}-k+i\Gamma }]  \notag \\
&&+(p_{1}\leftrightarrow p_{2})
\end{eqnarray}%
of two right-moving photons with momenta $p_{1}$ and $p_{2}$. The Fourier
transform of $S_{p_{1}p_{2},k}$ results in the wavefunction%
\begin{eqnarray}
\psi (x_{1},x_{2}) &=&\int \frac{dp_{1}dp_{2}}{2\pi }%
S_{p_{1}p_{2},k}e^{ip_{1}x_{1}+ip_{2}x_{2}}  \label{Ast} \\
&=&\int dxf_{\mathrm{in}}(x)[\mathcal{B}(x_{1},x_{2};x)+\mathcal{B}%
(x_{2},x_{1};x)],  \notag
\end{eqnarray}%
in the coordinate space, where%
\begin{eqnarray}
\mathcal{B}(x_{1},x_{2};x) &=&-i\sqrt{\Gamma }e^{\Gamma x_{2}}\theta
(-x_{2})[\delta (x-x_{1})  \notag \\
&&-\Gamma e^{-\Gamma (x-x_{1})}\theta (x_{2}-x)\theta (x-x_{1})],  \label{B}
\end{eqnarray}%
and%
\begin{equation}
f_{\mathrm{in}}(x)=\int \frac{dk}{\sqrt{2\pi }}f_{\mathrm{in}}(k)e^{ikx}
\end{equation}%
describes the incident wavepacket in the coordinate space.

We design the shape of the wavepacket $f_{\mathrm{in}}(x)$ to maximize the
stimulate emission probability%
\begin{equation}
P_{\mathrm{st}}=\frac{1}{2}\int dx_{1}dx_{2}\left\vert \psi
(x_{1},x_{2})\right\vert ^{2}.
\end{equation}%
The amplitude (\ref{Ast}) leads to%
\begin{equation}
P_{\mathrm{st}}=\int dxdyf_{\mathrm{in}}^{\ast }(x)W(x,y)f_{\mathrm{in}}(y),
\end{equation}%
where%
\begin{eqnarray}
W(x,y) &=&\int dx_{1}dx_{2}\mathcal{B}^{\ast }(x_{1},x_{2};x)\times  \notag
\\
&&[\mathcal{B}(x_{1},x_{2};y)+\mathcal{B}(x_{2},x_{1};y)].
\end{eqnarray}%
The function (\ref{B}) gives%
\begin{equation*}
W(x,y)=\frac{1}{2}\delta (y-x)+\frac{\Gamma }{4}(3e^{\Gamma
(x+y)}-e^{-\Gamma \left\vert x-y\right\vert })\theta (-x)\theta (-y).
\end{equation*}

The manifest effect of two photons to the system is realized by the initial
wavepacket $f_{\mathrm{in}}(x)$ localized at $x<0$. The largest eigenvalue
of $W(x,y)$ gives the maximal probability of stimulated emission, and the
corresponding eigenstate determines the shape of single photon wave-function
$f_{\mathrm{in}}(x)$.

Fortunately, the eigen-equation, i.e., the integral equation%
\begin{equation}
\int_{-\infty }^{0}dyW(x,y)f_{\mathrm{in}}(y)=\lambda f_{\mathrm{in}}(x)
\label{int}
\end{equation}%
can be solved exactly. In Eq. (\ref{int}), the second order derivative to $x$
leads to the differential equation%
\begin{equation}
\partial _{x}^{2}f_{\mathrm{in}}(x)=\frac{\lambda }{\lambda -\frac{1}{2}}%
\Gamma ^{2}f_{\mathrm{in}}(x),
\end{equation}%
where the solution is%
\begin{equation}
f_{\mathrm{in}}(x)=\left( \frac{4\Gamma ^{2}\lambda }{\lambda -\frac{1}{2}}%
\right) ^{1/4}\exp [\sqrt{\frac{\lambda }{\lambda -\frac{1}{2}}}\Gamma
x]\theta (-x).  \label{fin}
\end{equation}%
The eigenvalue $\lambda =2/3$ is obtained by the fact that $f_{\mathrm{in}%
}(x)$ is the solution of Eq. (\ref{int}). Finally, we conclude that the
largest probability of stimulated emission in the two-level emitter is $P_{%
\mathrm{st}}^{\max }=2/3$, and the corresponding incident wavepacket is $%
f(x)=2\sqrt{\Gamma }e^{2\Gamma x}\theta (-x)$.

We notice that this result agrees with that in Ref. \cite{Fanec}, where two
types of the incident wavepackets, i.e., the Gaussian type and the
Lorentzian shape are considered. By tuning the width of the wavepacket, the
maximal probability $2/3$ of the stimulated emission was found for the
initial wavepacket with the Lorentzian shape. Here, we proved the exact
result $P_{\mathrm{st}}^{\max }=2/3$ by solving the integral Eq. (\ref{int})
analytically.

\subsection{JC system}

In this section, we consider the JC system as the single emitter coupled to
waveguide photons, and apply the scattering theory to study the different
scattering processes. The emitter Hamiltonian in the rotating frame reads%
\begin{equation}
H_{\mathrm{sys}}=\Delta _{c}a^{\dagger }a+\Delta _{e}\left\vert
e\right\rangle \left\langle e\right\vert +g(a^{\dagger }\left\vert
g\right\rangle \left\langle e\right\vert +\mathrm{H.c.}),
\end{equation}%
where the detunings are $\Delta _{c}=\omega _{c}-k_{0}$ and $\Delta
_{e}=\omega _{e}-k_{0}$. For convenient, we focus on the resonant case $%
\Delta _{c}=\Delta _{e}=0$. The effective Hamiltonian $H_{\mathrm{eff}}=H_{%
\mathrm{sys}}-i\Gamma a^{\dagger }a$ follows from Eq. (\ref{Hes}). The
external source term is%
\begin{eqnarray}
S_{J} &=&-\sqrt{\frac{\Gamma }{L}}\int_{t_{i}}^{t_{f}}dt\sum_{k,\alpha
=r,l}[J_{k,\alpha }^{\ast }a(t)e^{-i\sigma _{\alpha }k(t_{f}-t)}  \notag \\
&&+J_{k,\alpha }a^{\dagger }(t)e^{-i\sigma _{\alpha }k(t-t_{i})}].
\end{eqnarray}

For the single incident photon with the momentum $k$ and the emitter
initially in the ground state, the boundary condition is $\gamma _{\mathrm{in%
}}=I$ and $\mathcal{F}_{\mathrm{in}}=\lim_{\{J_{k}\}\rightarrow 0}\delta
/\delta J_{k,r}$ in the asymptotic limit $t_{i}\rightarrow -\infty $. By the
boundary condition $\gamma _{\mathrm{out}}=I$ and $\mathcal{F}_{\mathrm{out}%
}=\lim_{\{J_{k}\}\rightarrow 0}\delta /\delta J_{p,(r,l)}$ in the asymptotic
limit $t_{f}\rightarrow \infty $, Eqs. (\ref{At}) and (\ref{AJt}) lead to
the reflection and transmission coefficients%
\begin{eqnarray}
R_{k}\delta _{p,-k} &=&-\frac{i\Gamma k}{(k+i\Gamma )k-g^{2}}\delta _{p,-k},
\notag \\
T_{k}\delta _{p,k} &=&\frac{k^{2}-g^{2}}{(k+i\Gamma )k-g^{2}}\delta _{p,k}.
\end{eqnarray}

For the two incident photons with momenta $k_{1}$ and $k_{2}$ to the emitter
initially in the ground state, the boundary condition is $\gamma _{\mathrm{in%
}}=I$ and $\mathcal{F}_{\mathrm{in}}=\lim_{\{J_{k}\}\rightarrow 0}\delta
^{2}/\delta J_{k_{1},r}\delta J_{k_{2},r}$ in the asymptotic limit $%
t_{i}\rightarrow -\infty $. By the boundary condition $\gamma _{\mathrm{out}%
}=I$ and $\lim_{\{J_{k}\}\rightarrow 0}\delta ^{2}/\delta J_{-p_{1},l}\delta
J_{-p_{2},l}$ in the asymptotic limit $t_{f}\rightarrow \infty $, Eqs. (\ref%
{At}) and (\ref{AJt}) lead to the $S$-matrix element%
\begin{eqnarray}
S_{p_{1}p_{2};k_{1}k_{2}} &=&\Gamma ^{2}\int_{-\infty }^{+\infty }\frac{%
dt_{1}^{\prime }dt_{2}^{\prime }dt_{1}dt_{2}}{(2\pi )^{2}}%
e^{ip_{1}t_{1}^{\prime }+ip_{2}t_{2}^{\prime }-ik_{1}t_{1}-ik_{2}t_{2}}
\notag \\
&&\times \left\langle \mathcal{T}a(t_{1}^{\prime })a(t_{2}^{\prime
})a^{\dagger }(t_{1})a^{\dagger }(t_{2})\right\rangle ,  \label{SJC2}
\end{eqnarray}%
of two reflected photons with momenta $-p_{1}$ and $-p_{2}$. Here, the
correlator for operators $a(t)=e^{iH_{\mathrm{eff}}t}ae^{-iH_{\mathrm{eff}%
}t} $ and $a^{\dagger }(t)=e^{iH_{\mathrm{eff}}t}a^{\dagger }e^{-iH_{\mathrm{%
eff}}t}$ leads to the $S$-matrix element%
\begin{eqnarray}
&&S_{-p_{1}-p_{2};k_{1}k_{2}}=R_{k_{1}}R_{k_{2}}(\delta _{p_{1}k_{1}}\delta
_{k_{2}p_{2}}+\delta _{p_{2}k_{1}}\delta _{k_{2}p_{1}})  \notag \\
&&+i\frac{\Gamma ^{2}}{\pi }\frac{g^{4}(E+i\Gamma )\delta
_{p_{1}+p_{2},k_{1}+k_{2}}}{(E+i\Gamma )(E+2i\Gamma )-2g^{2}}\times  \notag
\\
&&\frac{E(E+2i\Gamma )-4g^{2}}{\prod_{i=1,2}[k_{i}(k_{i}+i\Gamma
)-g^{2}][p_{i}(p_{i}+i\Gamma )-g^{2}]}.  \label{SJC3}
\end{eqnarray}%
The Fourier transform of Eq. (\ref{SJC3}) leads to the wavefunction%
\begin{eqnarray}
\psi (x_{c},x) &=&\frac{1}{2}%
\sum_{p_{1}p_{2}}S_{p_{1}p_{2};k_{1}k_{2}}e^{ip_{1}x_{1}+ip_{2}x_{2}}  \notag
\\
&=&e^{iEx_{c}}\{R_{k_{1}}R_{k_{2}}\cos (kx)  \notag \\
&&-\frac{\Gamma ^{2}g^{4}}{\lambda _{+}-\lambda _{-}}\frac{\sum_{s=\pm
}s(E-2\lambda _{s})e^{i(\frac{E}{2}-\lambda _{-s})x}}{(E+i\Gamma
)(E+2i\Gamma )-2g^{2}}  \notag \\
&&\times \frac{1}{\prod_{i=1,2}[k_{i}(k_{i}+i\Gamma )-g^{2}]}\}
\end{eqnarray}%
of two photons, where $\lambda _{s}$ is the solution of $k^{2}+i\Gamma
k-g^{2}=0$. As we discussed in Sec. III, under the weak driving light with
frequency $\omega _{d}$, the quantum statistics of photons emitting from the
JC system can be described by the second order correlation function $%
g^{(2)}(x)=\left\vert \psi (x_{c},x)\right\vert ^{2}$, where $\psi (x_{c},x)$
is the two photon wavefunction for the incident photons with frequency $%
k_{1}=k_{2}=\omega _{d}-k_{0}$. The above results agree with those in Ref.
\cite{JCShi}.

As a summary, in this section, we use the single emitter case as the example
to show how our theory works in the few photon scattering process. All the
results agree with the previous studies, which justifies the validity of the
developed scattering theory.

\section{Validity of Markov approximation}

In this section, we use the simple example to study the condition of the
Markov approximation in the array of two-level emitters. As shown in Secs.
III and IV, the dynamics of the single emitter coupled to the waveguide with
the linear dispersion can be exactly described by the effective Hamiltonian (%
\ref{Hes}) and the generalized master equation (\ref{gr}). For the array of
emitters, the dynamics is exactly characterized by the effective
time-nonlocal action $S_{\mathrm{eff}}$, where the effective Hamiltonian (%
\ref{Hes}) and the generalized master equation (\ref{gr}) only describe the
emitter evolution under the Markov approximation.

In order to justify the validity of Markov approximation, we compare the
exact result given by the path integral approach and the approximate result
based on the quantum regression theorem. This comparison shows that the
exact result and Markovian limit coincide when the bandwidth of the dynamics
is sufficiently small compared to the distance between emitters.

We focus on the single- and two- photon scattering processes, where the
effective action%
\begin{eqnarray}
S_{\mathrm{eff}} &=&\int d\omega \sum_{ij}b_{i}^{\dagger }(\omega )[\omega I-%
\mathcal{H}_{0}(\omega )]_{ij}b_{j}(\omega )  \notag \\
&&-\frac{U_{0}}{2}\int dt\sum_{j}b_{j}^{\dagger }(t)b_{j}^{\dagger
}(t)b_{j}(t)b_{j}(t)
\end{eqnarray}%
of emitters is given by the matrix with elements $[\mathcal{H}_{0}(\omega
)]_{ij}=-i\Gamma _{f}\delta _{ij}-i\Gamma e^{i(k_{0}+\omega )\left\vert
x_{i}-x_{j}\right\vert }$. The driving action $S_{J}=-%
\int_{t_{i}}^{t_{f}}dtH_{\mathrm{d}}(t)$ is given by Eq. (\ref{Hd}) with $%
O_{k,\sigma _{\alpha }}=\sqrt{\Gamma /L}\sum_{j}e^{-i(k+\sigma _{\alpha
}k_{0})x_{j}}b_{j}$. The equations (\ref{At}) and (\ref{AJt}) lead to the
exact results of the transmission spectrum and the second order correlation
function by the effective action $S_{\mathrm{eff}}$ and $S_{J}$. Under the
Markov approximation, Eqs. (\ref{At}) and (\ref{AJt})\ give the approximate
results by the effective Hamiltonian $\mathcal{H}_{0}(\omega )\sim \mathcal{H%
}_{0}(0)\equiv \mathcal{H}_{0}^{\mathrm{M}}$ and $O_{k,\sigma _{\alpha
}}\sim O_{0,\sigma _{\alpha }}$.

\subsection{Single photon processes}

For the single incident photon with the momentum $k$ and the emitters
initially in the ground state, the boundary condition is $\gamma _{\mathrm{in%
}}=I$ and $\mathcal{F}_{\mathrm{in}}=\lim_{\{J_{k}\}\rightarrow 0}\delta
/\delta J_{k,r}$ in the asymptotic limit $t_{i}\rightarrow -\infty $. It
follows from Eqs. (\ref{At}) and (\ref{AJt}) that the reflection coefficient%
\begin{equation}
R_{k}=-i\Gamma \sum_{ij}[G(k)]_{ij}e^{i(k+k_{0})(x_{i}+x_{j})}  \label{Rk}
\end{equation}%
of the photon with momentum $p=-k$ is determined by the boundary condition $%
\gamma _{\mathrm{out}}=I$ and $\mathcal{F}_{\mathrm{out}}=\lim_{\{J_{k}\}%
\rightarrow 0}\delta /\delta J_{p,l}$ in the asymptotic limit $%
t_{f}\rightarrow \infty $, where the Green function matrix $G(k)=[k-\mathcal{%
H}_{0}(k)]^{-1}$. In the Markovian limit, the reflection coefficient%
\begin{equation}
R_{k}^{\mathrm{M}}=-i\Gamma \sum_{ij}[G^{\mathrm{M}%
}(k)]_{ij}e^{ik_{0}(x_{i}+x_{j})}.  \label{RMk}
\end{equation}%
is determined by the approximate Green function $G^{\mathrm{M}}(k)=[k-%
\mathcal{H}_{0}(0)]^{-1}$.

\begin{figure}[tbp]
\includegraphics[bb=31 507 556 739, width=8 cm, clip]{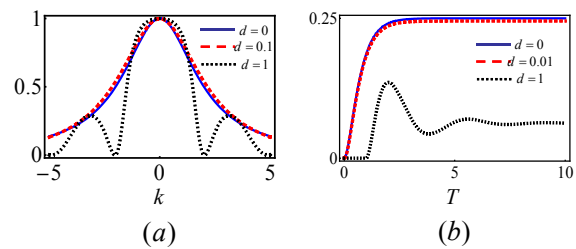}
\caption{(Color Online) (a) Single incident photon reflection spectra in a
two-emitter system, where the distance between emitters is $d$; (b)
Excitation probability of second emitter $|A(2,T)|^2$ as a function of time $%
T$ in a two-emitter system, when the second (first) emitter is initially
prepared in the ground (excited) state. Here, $\Gamma _{f}=0$ and $\Gamma $
is taken as the unit. The Markovian results are also given by the solid
(blue) curves.}
\label{SingM}
\end{figure}

The difference of Eqs. (\ref{Rk}) and (\ref{RMk}) is that the phase factor $%
e^{ikx_{i}}$ in $\mathcal{H}_{0}(k)$ is neglected in $G^{\mathrm{M}}(k)$,
which results in the condition $kNd\ll 1$ of the Markov approximation in the
single photon scattering. In Fig. \ref{SingM}a, we show the reflection
probabilities $\left\vert R_{k}\right\vert ^{2}$ and $\left\vert R_{k}^{%
\mathrm{M}}\right\vert ^{2}$ of the scattering photon by two atoms with
lattice spacing $d$, where $k_{0}d=2\pi n$ and $n$ is an integer. The
approximate result $\left\vert R_{k}^{\mathrm{M}}\right\vert ^{2}$ does not
depend on $d$, which agrees with the exact result $\left\vert
R_{k}\right\vert ^{2}$ very well for small $kd\ll 1$. In the non-Markovian
regime $kd>1$, the dotted (black) curve ($d=1$) shows that the reflection
probability has the peaks localized around $n_{0}\pi /d$ and $n_{0}$ is an
integer. The positions of these peaks are the resonant frequencies of
eigenmodes in the \textquotedblleft cavity\textquotedblright\ formed by the
two emitters.

The behavior of single excitation propagation in the emitter array can also
be analyzed by Eqs. (\ref{At}) and (\ref{AJt}), where the exact result gives
the condition of the Markov approximation. For the first emitter initially
in the excited state and the rest emitters in the ground state, the boundary
conditions are $\mathcal{F}_{\mathrm{in}}=I$ and $\gamma _{\mathrm{in}%
}=b_{1} $ at the initial instant $t_{i}=0$. The propagation of single
excitation in the emitter array is described by the boundary condition $%
\mathcal{F}_{\mathrm{out}}=I$ and $\gamma _{\mathrm{out}}=b_{j}$ at the
final instant $t_{f}=T$. The equations (\ref{At}) and (\ref{AJt}) lead to
the amplitude%
\begin{equation}
\mathcal{A}(j,T)=i\int \frac{d\omega }{2\pi }[\frac{1}{\omega -\mathcal{H}%
_{0}(\omega )}]_{j1}e^{-i\omega T}  \label{A2L1}
\end{equation}%
of detecting the excitation at the $j$-th emitter. The single photon emitted
by the excitation of the first emitter is described by the amplitude%
\begin{eqnarray}
&&\mathcal{A}_{\alpha }(x,T)=-i\sqrt{\Gamma }\sum_{j}e^{-i\sigma _{\alpha
}k_{0}x_{j}}\theta (\sigma _{\alpha }(x-x_{j}))  \notag \\
&&\times \theta (T-\sigma _{\alpha }(x-x_{j}))\mathcal{A}(j,T-\sigma
_{\alpha }(x-x_{j}))  \label{A2L2}
\end{eqnarray}%
of detecting a single photon at the position $x$ in the waveguide, where $%
\alpha $ denotes the right- and left- moving modes, and the boundary
condition is $\mathcal{F}_{\mathrm{out}}=\lim_{\{J_{k}\}\rightarrow 0}\delta
/\delta J_{k,(r,l)}$ and $\gamma _{\mathrm{out}}=I$ at the final instant $%
t_{f}=T$. In the Markovian limit, the amplitudes become%
\begin{equation}
\mathcal{A}^{\mathrm{M}}(j,T)=i\int \frac{d\omega }{2\pi }[\frac{1}{\omega -%
\mathcal{H}_{0}(0)}]_{j1}e^{-i\omega T},
\end{equation}%
and%
\begin{eqnarray}
&&\mathcal{A}_{\alpha }^{\mathrm{M}}(x,T)=-i\sqrt{\Gamma }%
\sum_{j}e^{-i\sigma _{\alpha }k_{0}x_{j}}\theta (\sigma _{\alpha }(x-x_{j}))
\notag \\
&&\times \theta (T-\sigma _{\alpha }(x-x_{j}))\mathcal{A}^{\mathrm{M}%
}(j,T-\sigma _{\alpha }(x-x_{j})).
\end{eqnarray}

To investigate the non-Markov effects and justify the validity of the Markov
approximation, we compare the exact results and the approximate results. It
follows from the residue theorem that the poles $\xi _{\lambda }$ and the
corresponding residues $Z_{\lambda }$ of $[1/(\omega -\mathcal{H}_{0}(\omega
))]_{j1}$ determine the time evolution $\mathcal{A}(j,T)\sim \sum_{\lambda
}Z_{\lambda }e^{-i\xi _{\lambda }T}$. Similarly, $\mathcal{A}^{\mathrm{M}%
}(j,T)\sim \sum_{\lambda }Z_{\lambda }^{\mathrm{M}}e^{-i\xi _{\lambda }^{%
\mathrm{M}}T}$ is given by the poles $\xi _{\lambda }^{\mathrm{M}}$ and the
corresponding residues $Z_{\lambda }^{\mathrm{M}}$ of $[1/(\omega -\mathcal{H%
}_{0}(0))]_{j1}$. Under the condition $\xi _{\lambda }^{\mathrm{M}}Nd\ll 1$,
the phase factor $e^{i\xi _{\lambda }^{\mathrm{M}}\left\vert
x_{i}-x_{j}\right\vert }\sim 1$, and $\xi _{\lambda }^{\mathrm{M}}$ and $%
Z_{\lambda }^{\mathrm{M}}$ are approximately the pole $\xi _{\lambda }$ and
the corresponding residue $Z_{\lambda }$ of the exact propagator $[1/(\omega
-\mathcal{H}_{0}(\omega ))]_{j1}$. As a result, the condition of the Markov
approximation is $\xi _{\lambda }^{\mathrm{M}}Nd\ll 1$.
\begin{figure}[tbp]
\includegraphics[bb=21 198 575 764, width=8 cm, clip]{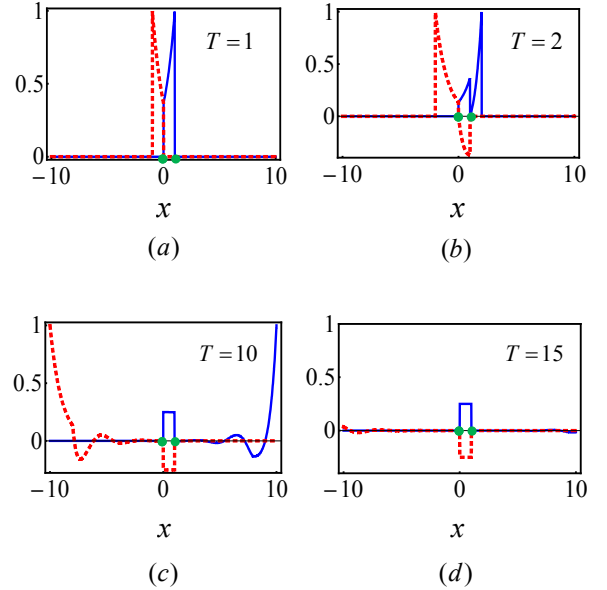}
\caption{(Color Online) The solid (blue) and the dashed (red) curves show
the wavefunctions of the right- and left- moving photon after the evolution
time $T$, where the first (second) emitter is initially prepared in the
excited (ground) state. Here, $\Gamma _{f}=0$ and $\Gamma$ is taken as the
unit, and the green dots denote the two emitters.}
\label{singpro2}
\end{figure}

To understand this condition, we consider two emitters in the waveguide,
where the first emitter is placed at the origin, and the effective action is
determined by the $2\times 2$ matrix%
\begin{equation}
\mathcal{H}_{0}(\omega )=\left(
\begin{array}{cc}
-i\Gamma _{f}-i\Gamma & -i\Gamma e^{i(k_{0}+\omega )d} \\
-i\Gamma e^{i(k_{0}+\omega )d} & -i\Gamma _{f}-i\Gamma%
\end{array}%
\right) .
\end{equation}%
It follows from Eqs. (\ref{A2L1}) and (\ref{A2L2}) that the exact amplitudes
are%
\begin{eqnarray}
\mathcal{A}(1,T) &=&\frac{1}{2}e^{-(\Gamma _{f}+\Gamma
)T}[C_{+}(T)+C_{-}(T)],  \notag \\
\mathcal{A}(2,T) &=&\frac{1}{2}e^{-(\Gamma _{f}+\Gamma
)T}[C_{-}(T)-C_{+}(T)],  \label{AAC}
\end{eqnarray}%
and%
\begin{eqnarray}
\mathcal{A}_{\alpha }(x,T) &=&-i\sqrt{\Gamma }[\theta (\sigma _{\alpha
}x)\theta (T-\sigma _{\alpha }x)\mathcal{A}(1,T-\sigma _{\alpha }x)  \notag
\\
&&+e^{-i\sigma _{\alpha }k_{0}d}\theta (\sigma _{\alpha }(x-d))\theta
(T-\sigma _{\alpha }(x-d))  \notag \\
&&\times \mathcal{A}(2,T-\sigma _{\alpha }x+\sigma _{\alpha }d)],  \label{A2}
\end{eqnarray}%
where%
\begin{equation}
C_{\pm }(T)=\sum_{n=0}^{\infty }\frac{1}{n!}[\pm \Gamma e^{ik_{0}d+(\Gamma
_{f}+\Gamma )d}(T-nd)]^{n}\theta (T-nd).  \label{Cpm}
\end{equation}%
The amplitude $\mathcal{A}_{\alpha }^{\mathrm{M}}(x,T)$ under the Markov
approximation has the same form of Eq. (\ref{A2}), where $\mathcal{A}(j,T)$
is approximated by%
\begin{eqnarray}
\mathcal{A}^{\mathrm{M}}(1,T) &=&\cosh (\Gamma Te^{ik_{0}d})e^{-(\Gamma
_{f}+\Gamma )T},  \notag \\
\mathcal{A}^{\mathrm{M}}(2,T) &=&-\sinh (\Gamma Te^{ik_{0}d})e^{-(\Gamma
_{f}+\Gamma )T}.  \label{AM}
\end{eqnarray}%
The agreement between Eq. (\ref{AM}) and the exact result (\ref{AAC}) is
verified in the Markovian limit $(\Gamma _{f}+\Gamma )d\ll 1$, where $C_{\pm
}(T)\sim e^{\pm \Gamma Te^{ik_{0}d}}$.

The time evolution of the two emitters exhibits two non-Markovian effects.
The first is the retardation effect. The amplitude $\mathcal{A}^{\mathrm{M}%
}(2,T)$ shows that once $T>0$ the second emitter has the probability in the
excited state. However, this is an artificial effect of the instantaneous
effective Hamiltonian $H_{\mathrm{eff}}$ under the Markov approximation. In
fact, the single photon wavepacket emitted by the first emitter takes the
time $T=d$ to arrive at the second emitter. Hence, the second emitter has to
stay at the ground state for $T<d$, i.e., $\mathcal{A}(2,T<d)=0$. This
retardation effect is fully characterized by the exact result (\ref{AAC})
and $C_{\pm }(T)$. We show the probability $\left\vert \mathcal{A}%
(2,T)\right\vert ^{2}$ in Fig. \ref{SingM}b for $\Gamma _{f}=0$ and $%
k_{0}d=2\pi n$, where the dotted (black) curve for $d=1$ displays the
retardation effect explicitly. In the small $d$ limit, i.e., $\Gamma d/c\ll
1 $, the Markovian result agrees with the exact result very well, as shown
by the solid (blue) and dashed (red) curves in Fig. \ref{SingM}b, where the
speed of light $c$ is taken to be the unit.

The second effect is the formation of the entangle state $(\left\vert
eg\right\rangle -\left\vert ge\right\rangle )/\sqrt{2}$ in the limit $%
T\rightarrow \infty $. For the vanishing $\Gamma _{f}=0$, both the exact and
Markovian results show that in the steady state $T\rightarrow \infty $ the
emitters have the probability $P_{e}$ to form the entangle state. Under the
Markov approximation, the probability $P_{e}=1/2$ is determined by $\mathcal{%
A}^{\mathrm{M}}(1,\infty )=-\mathcal{A}^{\mathrm{M}}(2,\infty )=1/2$. The
exact result $\mathcal{A}(1,\infty )=-\mathcal{A}(2,\infty )=1/(2+2\Gamma d)$
gives the probability $P_{e}=1/[2(1+\Gamma d)^{2}]$ to form the entangle
state. In the limit $\Gamma d\ll 1$, the Markov approximation works
perfectly, i.e., $\mathcal{A}(j,\infty )\sim \mathcal{A}^{\mathrm{M}%
}(j,\infty )=1/2$.

During the formation of the entangled state, the dynamics of the waveguide
photon is described by the amplitudes $\mathcal{A}_{r,l}(x,T)$. In Fig. \ref%
{singpro2}, we show the propagation of the right- and left- moving
wavepackets by $\mathcal{A}_{r,l}(x,T)$ for the distance $d=1$. In the
steady state $T\rightarrow \infty $, the entangle state is established,
where the standing wave in the regime $0<x<d$ forms to mediate the
interaction of two emitters.

\subsection{Two photon processes}

In this section, we study the scattering process of two incident photons
with momenta $k_{1}$ and $k_{2}$. By comparing the exact result and the
approximate result, we investigate the condition of the Markov approximation
in the two-photon processes.

\begin{figure}[tbp]
\includegraphics[bb=21 283 583 769, width=8 cm, clip]{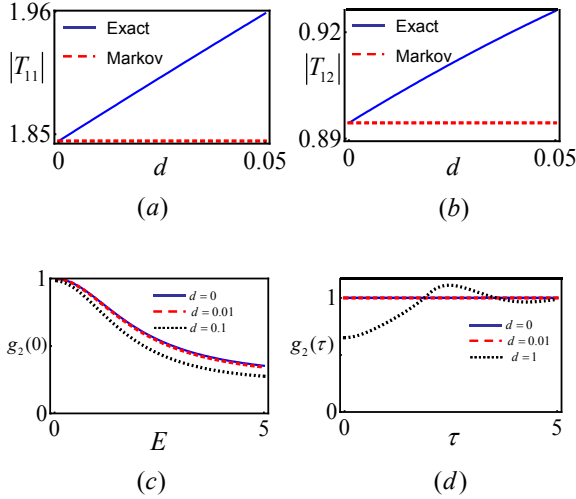}
\caption{(Color Online) (a)-(b) The exact and Markovian T-matrix elements
for $E=1$; (c)-(d) The second order correlation functions of out-going
photons: In (c) $\protect\tau=0$, in (d) $E=0$, and Markovian results are
shown by the solid (blue) curves. Here, $\Gamma_{f}=0$ and $\Gamma$ is taken
as the unit.}
\label{T22}
\end{figure}

The initial boundary condition is $\gamma _{\mathrm{in}}=I$ and $\mathcal{F}%
_{\mathrm{in}}=\lim_{\{J_{k}\}\rightarrow 0}\delta ^{2}/\delta
J_{k_{1},r}\delta J_{k_{2},r}$ in the asymptotic limit $t_{i}\rightarrow
-\infty $. It follows from Eqs. (\ref{At}) and (\ref{AJt}) that the $S$%
-matrix%
\begin{eqnarray}
&&S_{p_{1}p_{2},k_{1}k_{2}}=R_{k_{1}}R_{k_{2}}(\delta _{p_{1}k_{1}}\delta
_{p_{2}k_{2}}+\delta _{p_{2}k_{1}}\delta _{p_{1}k_{2}})  \label{S22} \\
&&-i\frac{\Gamma ^{2}}{\pi }\delta _{p_{1}+p_{2},k_{1}+k_{2}}\sum_{ij}\bar{G}%
_{i}(p_{1},p_{2})T_{ij}(E)\bar{G}_{j}(k_{1},k_{2})  \notag
\end{eqnarray}%
of two reflected photons with momenta $-p_{1}$ and $-p_{2}$ are determined
by the boundary condition $\gamma _{\mathrm{out}}=I$ and $\mathcal{F}_{%
\mathrm{out}}=\lim_{\{J_{k}\}\rightarrow 0}\delta ^{2}/\delta
J_{-p_{1},l}\delta J_{-p_{2},l}$ in the asymptotic limit $t_{f}\rightarrow
\infty $. Here,%
\begin{equation}
\bar{G}_{i}(p_{1},p_{2})=%
\sum_{i_{1}i_{2}}G_{i_{1}i}(p_{1})G_{i_{2}i}(p_{2})e^{i(p_{1}+k_{0})x_{i_{1}}}e^{i(p_{2}+k_{0})x_{i_{2}}},
\end{equation}%
and the $T$-matrix $T(E)=-\Pi ^{-1}(E)$ is given by the Dyson Expansion with
the bubble%
\begin{equation}
\Pi _{ij}(E)=i\int \frac{d\omega }{2\pi }G_{ij}(\omega )G_{ij}(E-\omega )
\label{bubble}
\end{equation}%
and $E=k_{1}+k_{2}$.

Under the Markov approximation, $S_{p_{1}p_{2},k_{1}k_{2}}$ is given by the
same form of Eq. (\ref{S22}), where $R_{k_{i}}\sim R_{k_{i}}^{\mathrm{M}}$,%
\begin{equation}
\bar{G}_{i}(p_{1},p_{2})\sim \sum_{i_{1}i_{2}}G_{i_{1}i}^{\mathrm{M}%
}(p_{1})G_{i_{2}i}^{\mathrm{M}}(p_{2})e^{ik_{0}(x_{i_{1}}+x_{i_{2}})}
\end{equation}%
and the bubble (\ref{bubble}) are determined by the approximate Green
function $G^{\mathrm{M}}(\omega )\sim \lbrack \omega -\mathcal{H}%
_{0}(0)]^{-1}$. The exact result shows that the Markov approximation is
valid in the limit $k_{i}Nd$, $p_{i}Nd$, and $\Gamma Nd\ll 1$.

In order to understand the condition, we investigate the two-photon
scattering processes explicitly by considering the photon scattering by two
emitters, where $k_{0}d=2\pi n$. We first compare the exact $T$-matrix $T(E)$
with the $T$-matrix $T^{\mathrm{M}}(E)$ under the Markov approximation,
where the elements $\Pi _{11}^{\mathrm{M}}(E)=\Pi _{22}^{\mathrm{M}}(E)=\Pi
_{a}^{\mathrm{M}}(E)+\Pi _{b}^{\mathrm{M}}(E)$ and $\Pi _{12}^{\mathrm{M}%
}(E)=\Pi _{21}^{\mathrm{M}}(E)=\Pi _{a}^{\mathrm{M}}(E)-\Pi _{b}^{\mathrm{M}%
}(E)$ are given by%
\begin{eqnarray}
\Pi _{a}^{\mathrm{M}}(E) &=&\frac{1}{4}\sum_{\sigma =\pm 1}\frac{1}{%
E+2i\Gamma _{f}+2i\Gamma (1+\sigma e^{ik_{0}d})},  \notag \\
\Pi _{b}^{\mathrm{M}}(E) &=&\frac{1}{2}\frac{1}{E+2i\Gamma _{f}+2i\Gamma }.
\end{eqnarray}%
The absolute values of the exact $T$-matrix elements are shown in Figs. \ref%
{T22}a and \ref{T22}b for $k_{0}d=2\pi n$, which illustrate the perfect
agreement between the Markovian result $T^{\mathrm{M}}(E)$ and the exact
result in the limit $Ed,\Gamma d\ll 1$.

The Fourier transformation of $S$-matrix leads to the wavefunction%
\begin{equation}
\psi (x_{c},x)=\frac{1}{2}\int
dp_{1}dp_{2}S_{p_{1}p_{2},k_{1}k_{2}}e^{ip_{1}x_{1}+ip_{2}x_{2}}
\end{equation}%
of two reflected photons, where the center of mass coordinate $%
x_{c}=(x_{1}+x_{2})/2$ and the relative coordinate $x=x_{1}-x_{2}$. For two
incident photons with the same momentum $k_{1}=k_{2}=k$, the photon
statistics is characterized by the second order correlation function%
\begin{equation}
g^{(2)}(\tau )=\frac{\left\vert \psi (x_{c},\tau )\right\vert ^{2}}{%
\left\vert R_{k}\right\vert ^{4}}.
\end{equation}

In Figs. \ref{T22}c and \ref{T22}d, we show $g^{(2)}(0)$ as the function of $%
E=2k$ and $g^{(2)}(\tau )$ for the resonant frequency $k=0$, where the
Markovian results perfectly agree with the exact result in the small $kd$
and $\Gamma d$ limit. When $d$ is increasing, e.g., $\Gamma d=0.1$ and $1$,
the Markovian result deviates from the exact one.

As the summary in this section, we use the exact result to study the
non-Markovian effects and examine the condition of the Markov approximation.
We conclude that in the limit $\Gamma Nd,kNd\ll 1$, the Markov approximation
works perfectly. For the large $\Gamma Nd$ and $kNd$, some non-Markov
effects emerge, e.g., the retardation effect. In the following sections, we
assume the system length $Nd$ is small enough and the Markov approximation
is always valid.

\section{Entangled photon pairs by single emitter}

In this section, we use the scattering theory to study the generation of
entangled photons in the scattering process. In order to realize the
deterministic generation of reflected photons, the two-level emitter is
placed at the left hand side of a perfect mirror, i.e., in the half-end
waveguide, as shown in Fig. \ref{fig1}c. Here, the mirror is put at the
origin, and the position of the emitter is $x_{0}<0$.

It follows from Eq. (\ref{re}) that the effective action is%
\begin{eqnarray}
S_{\mathrm{eff}} &=&S_{\mathrm{sys}}+i\int dt\Gamma e^{\dagger
}(t)e(t)+i\int dt\Gamma _{b}b^{\dagger }(t)b(t)  \notag \\
&&+i\sqrt{\Gamma \Gamma _{b}}\int dtb^{\dagger }(t)e(t+x_{0})e^{-ik_{0}x_{0}}
\notag \\
&&+i\sqrt{\Gamma \Gamma _{b}}\int dte^{\dagger
}(t)b(t+x_{0})e^{-ik_{0}x_{0}},
\end{eqnarray}%
where the emitter action $S_{\mathrm{sys}}$ is determined by the\
Hamiltonian $H_{\mathrm{sys}}=H_{\mathrm{emitter}}$. By the creation
(annihilation) operator $e^{\dagger }$ ($e$), the hardcore boson is
introduced to describe the two-level emitter, where the energy level spacing
is $\omega _{e}=k_{0}$ and $H_{\mathrm{emitter}}=U_{0}e^{\dagger }e^{\dagger
}ee/2$ in the limit $U_{0}\rightarrow \infty $. The external source term $%
S_{J}=-\int_{t_{i}}^{t_{f}}dtH_{\mathrm{d}}(t)$ is given by Eqs. (\ref{Hd})
and (\ref{O}) with the jump operator $O=e$.

\subsection{Single- and two- photon scattering}

For the single incident photon with the momentum $k$ and the emitter
initially in the ground state, the boundary condition is $\gamma _{\mathrm{in%
}}=I$ and $\mathcal{F}_{\mathrm{in}}=\lim_{\{J_{k}\}\rightarrow 0}\delta
/\delta J_{k,r}$ in the asymptotic limit $t_{i}\rightarrow -\infty $. By the
boundary condition $\gamma _{\mathrm{out}}=I$ and $\mathcal{F}_{\mathrm{out}%
}=\lim_{\{J_{k}\}\rightarrow 0}\delta /\delta J_{p,l}$ in the asymptotic
limit $t_{f}\rightarrow \infty $, Eq. (\ref{At}) and (\ref{AJt}) result in
the reflection coefficient%
\begin{eqnarray}
R_{k}\delta _{p,-k} &=&-i\delta _{p,-k}[\Gamma
G_{ee}(k)e^{2i(k+k_{0})x_{0}}+\Gamma _{b}G_{bb}(k)  \notag \\
&&+2\sqrt{\Gamma \Gamma _{b}}G_{eb}(k)e^{i(k+k_{0})x_{0}}],
\end{eqnarray}%
where the Green functions $G_{ss^{\prime }}(k)=-i\int dte^{ikt}\left\langle
s(t)s^{\prime \dagger }\right\rangle $ with $s,s^{\prime }=e,b$. The
effective action determines the single particle Green functions%
\begin{eqnarray}
&&\left(
\begin{array}{cc}
G_{ee}(k) & G_{eb}(k) \\
G_{be}(k) & G_{bb}(k)%
\end{array}%
\right)  \notag \\
&=&\left(
\begin{array}{cc}
k+i\Gamma & i\sqrt{\Gamma \Gamma _{b}}e^{-i(k_{0}+k)x_{0}} \\
i\sqrt{\Gamma \Gamma _{b}}e^{-i(k_{0}+k)x_{0}} & k+i\Gamma _{b}%
\end{array}%
\right) ^{-1},
\end{eqnarray}%
which give the reflection coefficient%
\begin{equation}
R_{k}\delta _{p,-k}=-\frac{k-i\Gamma +i\Gamma e^{-2i(k+k_{0})\left\vert
x_{0}\right\vert }}{k+i\Gamma -i\Gamma e^{2i(k_{0}+k)\left\vert
x_{0}\right\vert }}\delta _{p,-k}
\end{equation}%
and $\left\vert R_{k}\right\vert =1$ in the limit $\Gamma _{b}\rightarrow
\infty $. The phase shift arg$(R_{k})$ is shown in Figs. \ref{fig3}a for $%
d=10^{-4}$, where $\theta _{0}=k_{0}\left\vert x_{0}\right\vert $.

For two incident photons with momenta $k_{1}$ and $k_{2}$ and the emitter
initially in the ground state, the boundary condition is $\gamma _{\mathrm{in%
}}=I$ and $\mathcal{F}_{\mathrm{in}}=\lim_{\{J_{k}\}\rightarrow 0}\delta
^{2}/\delta J_{k_{1},r}\delta J_{k_{2},r}$ in the asymptotic limit $%
t_{i}\rightarrow -\infty $. By the boundary condition $\gamma _{\mathrm{out}%
}=I$ and $\mathcal{F}_{\mathrm{out}}=\lim_{\{J_{k}\}\rightarrow 0}\delta
^{2}/\delta J_{-p_{1},l}\delta J_{-p_{2},l}$ in the asymptotic limit $%
t_{i}\rightarrow -\infty $, the $S$-matrix element%
\begin{eqnarray}
&&S_{-p_{1}-p_{2};k_{1}k_{2}}=R_{k_{1}}R_{k_{2}}(\delta _{p_{1}k_{1}}\delta
_{p_{2}k_{2}}+\delta _{p_{1}k_{2}}\delta _{p_{1}k_{2}})  \notag \\
&&-i\frac{16}{\pi }\Gamma ^{2}\delta _{p_{1}+p_{2},k_{1}+k_{2}}T(E)\times
\label{S3} \\
&&\prod_{i=1,2}\frac{\sin [(k_{i}+k_{0})\left\vert x_{0}\right\vert ]\sin
[(p_{i}+k_{0})\left\vert x_{0}\right\vert ]}{[k_{i}+i\Gamma -i\Gamma
e^{2i(k_{0}+k_{i})\left\vert x_{0}\right\vert }][p_{i}+i\Gamma -i\Gamma
e^{2i(k_{0}+p_{i})\left\vert x_{0}\right\vert }]}  \notag
\end{eqnarray}%
of two reflected photons with momenta $p_{1}$ and $p_{2}$ is given by Eqs. (%
\ref{At}) and (\ref{AJt}) in the asymptotic limit $t_{f}\rightarrow \infty $%
. Here, the $T$-matrix element%
\begin{equation}
T(E)=\frac{1}{U_{0}^{-1}-\Pi (E)}
\end{equation}%
is determined by the vacuum bubble%
\begin{equation}
\Pi (E)=i\int \frac{d\omega }{2\pi }G_{ee}(\omega )G_{ee}(E-\omega )
\end{equation}%
and $E=k_{1}+k_{2}=p_{1}+p_{2}$.

\subsection{Entangled photon pairs}

\begin{figure}[tbp]
\includegraphics[bb=17 469 571 731, width=8 cm, clip]{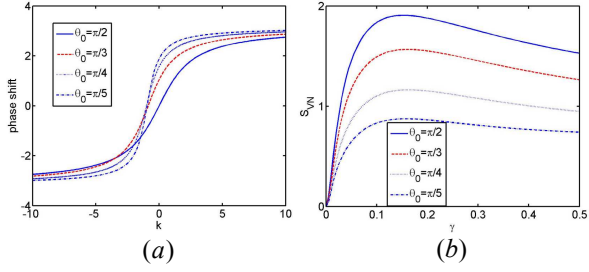}
\caption{(Color Online) (a) The phase shift of the single photon; (b) The
Von-Neumann entropy of the reflective photons. Here, $d=10^{-4}$, and $%
\Gamma $ is taken as the unit.}
\label{fig3}
\end{figure}

For the state $\sum_{k_{1}k_{2}}f(k_{1})f(k_{2})r_{k_{1}}^{\dagger
}r_{k_{2}}^{\dagger }\left\vert 0\right\rangle /\sqrt{2}$ of two independent
incident photons, the $S$-matrix (\ref{S3}) leads to the asymptotic state%
\begin{equation}
\left\vert \psi _{\mathrm{out}}\right\rangle =\frac{1}{\sqrt{2}}%
\sum_{p_{1}p_{2}}\psi _{\mathrm{out}}(p_{1},p_{2})l_{-p_{1}}^{\dagger
}l_{-p_{2}}^{\dagger }\left\vert 0\right\rangle
\end{equation}%
of two reflected photons, where the wavefunction is%
\begin{eqnarray}
\psi _{\mathrm{out}}(p_{1},p_{2})
&=&f(p_{1})f(p_{2})R_{p_{1}}R_{p_{2}}-16i\Gamma ^{2}T(E)F_{2}(E)  \notag \\
&&\times \prod_{i=1,2}\frac{\sin [(p_{i}+k_{0})\left\vert x_{0}\right\vert ]%
}{p_{i}+i\Gamma -i\Gamma e^{2i(k_{0}+p_{i})\left\vert x_{0}\right\vert }},
\end{eqnarray}%
and the integral%
\begin{equation}
F_{2}(E)=\int \frac{dk}{2\pi }\prod_{\sigma =\pm 1}\frac{f(\frac{E}{2}%
+\sigma k)\sin [(\frac{E}{2}+\sigma k+k_{0})\left\vert x_{0}\right\vert ]}{%
\frac{E}{2}+\sigma k+i\Gamma -i\Gamma e^{2i(\frac{E}{2}+\sigma
k+k_{0})\left\vert x_{0}\right\vert }}.
\end{equation}

The Von-Neumann entropy $S_{\mathrm{VN}}=-tr(\rho _{1}\ln \rho _{1})$ is
given by the single-photon reduced density matrix $\rho
_{1}=tr_{2}(\left\vert \psi _{\mathrm{out}}\right\rangle \left\langle \psi _{%
\mathrm{out}}\right\vert )$, where the degree of freedom for the other
photon is traced out. The Von-Neumann entropy $S_{\mathrm{VN}%
}=-\sum_{\lambda }\lambda ^{2}\ln \lambda ^{2}$ can be obtained by the
singular value decomposition of $\psi _{\mathrm{out}}(p_{1},p_{2})=\sum_{%
\lambda }g_{\lambda }(p_{1})\lambda \tilde{g}_{\lambda }(p_{2})$. Here, the
singluar values $\lambda $ measure the entanglement of two photons in
different modes $g_{\lambda }$ and $\tilde{g}_{\lambda }$. For the single
photon wavepacket%
\begin{equation}
f(k)=\sqrt{\frac{\gamma }{\pi }}\frac{1}{k+i\gamma },
\end{equation}%
we show $S_{\mathrm{VN}}$ in Fig. \ref{fig3}b for $d=10^{-4}$, which
displays the generation of entangled photons by the two-level emitter.

\section{Rydberg-EIT system}

In this section, we consider the application of the scattering theory in the
photon transmission in the EIT atoms coupled to the Rydberg level. We shall
show that the developed scattering theory is an efficient approach to the
photon transmission in the array of interacting emitters with complicated
structures. Here, we highlight the interplay between the EIT phenomenon and
the Rydberg interaction, and show a rich variety of quantum statistics of
the scattering photons and polariton exitations.

In the rotating frame, the Hamiltonian (\ref{HR}) becomes%
\begin{eqnarray}
H_{\mathrm{sys}} &=&\sum_{i}[\Delta _{e}e_{i}^{\dagger }e_{i}+\Delta
_{s}s_{i}^{\dagger }s_{i}+(\Omega e_{i}^{\dagger }s_{i}+\mathrm{H.c.})]
\notag \\
&&+H_{\mathrm{HC}}+\frac{1}{2}\sum_{ij}U_{ij}s_{i}^{\dagger }s_{j}^{\dagger
}s_{j}s_{i}.
\end{eqnarray}%
where $\Delta _{e}=\omega _{e}-k_{0}$, $\Delta _{s}=\omega _{s}-k_{0}+\omega
_{d}$, and we focus on the two photon resonance case $\Delta _{s}=0$. The
waveguide photons couple to the $N$ atoms collectively through the operator $%
O_{k,\pm }=\sqrt{\Gamma }\sum_{i=1}^{N}e^{-i(k\pm k_{0})x_{i}}e_{i}/\sqrt{L}$%
, and the free space modes couple to the atom operator $e_{i}$ locally,
which induces the decay of the excited state $e$. The effective action%
\begin{eqnarray}
S_{\mathrm{eff}} &=&\int dt\{\sum_{i}[e_{i}^{\dagger }(t)(i\partial
_{t}+i\Gamma _{f})e_{i}(t)  \notag \\
&&+s_{i}^{\dagger }(t)i\partial _{t}s_{i}(t)]-H_{\mathrm{sys}}\}+S_{\mathrm{%
re}},
\end{eqnarray}%
of EIT atoms is given by%
\begin{equation}
S_{\mathrm{re}}=i\Gamma \int dt\sum_{ij}e_{i}^{\ast }(t)e_{j}(t-\left\vert
x_{i}-x_{j}\right\vert )e^{ik_{0}\left\vert x_{i}-x_{j}\right\vert }.
\end{equation}%
The source term $S_{J}=-\int_{t_{i}}^{t_{f}}dtH_{\mathrm{d}}(t)$ is
determined by Eq. (\ref{Hd}). Here, the lattice spacing $d$ satisfies $%
k_{0}d=(2n+1/2)\pi $.

Based on the general result (\ref{At}) and (\ref{AJt}), we first study the
single- and two- photon scattering, and show the transmission spectrum and
the second order correlation function of out-going photons. In the second
part, we investigate the transient processes and show how the wavepackets of
single and two incident photons transfer to the atom excitations, propagate
in the Rydberg-EIT atom array, and finally emit back to the waveguide. For
the single incident photon, we show the free propagation of the dark
polariton. For two incident photons, we show the counter-propagation and
co-propagation of two polaritons.

\subsection{Single and two photon scatterings}

\begin{figure}[tbp]
\includegraphics[bb=18 247 563 723, width=8 cm, clip]{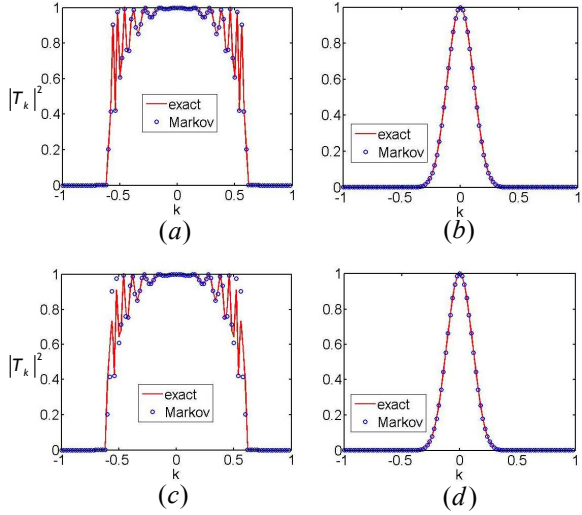}
\caption{(Color Online) The single photon transmission spectra, where $%
\Gamma=1$, $\Delta_e=0$, $\Omega=1$, and the atom number is 20. (a) $%
\Gamma_f =0$ and $d=10^{-4}$; (b) $\Gamma_f=1$ and $d=10^{-4}$; (c) $%
\Gamma_f =0$ and $d=10^{-2}$; (d) $\Gamma_f=1$ and $d=10^{-2}$.}
\label{singlepolaron}
\end{figure}

For the single incident photon with the momentum $k$ and the atoms initially
in the ground state, the boundary condition is $\gamma _{\mathrm{in}}=I$ and
$\mathcal{F}_{\mathrm{in}}=\lim_{\{J_{k}\}\rightarrow 0}\delta /\delta
J_{k,r}$ in the asymptotic limit $t_{i}\rightarrow -\infty $. By the
boundary condition $\gamma _{\mathrm{out}}=I$ and $\mathcal{F}_{\mathrm{out}%
}=\lim_{\{J_{k}\}\rightarrow 0}\delta /\delta J_{p,(r,l)}$ in the asymptotic
limit $t_{f}\rightarrow \infty $, Eqs. (\ref{At}) and (\ref{AJt}) lead to
the reflection and transmission coefficients%
\begin{eqnarray}
R_{k}\delta _{p,-k} &=&-i\Gamma \delta
_{p,-k}\sum_{ij}[G_{0}(k)]_{ij}^{ee}e^{i(k+k_{0})(x_{i}+x_{j})},  \label{RT}
\\
T_{k}\delta _{p,k} &=&\delta _{pk}\{1-i\Gamma
\sum_{ij}[G_{0}(k)]_{ij}^{ee}e^{-i(k+k_{0})(x_{i}-x_{j})}\},  \notag
\end{eqnarray}%
of the photon with momentum $p$.

The free Green function%
\begin{equation}
G_{0}(\omega )=\frac{1}{\omega -\mathcal{H}_{0}(\omega )}
\end{equation}%
of atoms is determined by the block form%
\begin{equation}
\mathcal{H}_{0}(\omega )=\left(
\begin{array}{cc}
(\Delta _{e}-i\Gamma _{f})\delta _{ij}-i\Gamma e^{i(k_{0}+\omega )\left\vert
x_{i}-x_{j}\right\vert } & \Omega \delta _{ij} \\
\Omega \delta _{ij} & \Delta _{s}\delta _{ij}%
\end{array}%
\right)  \label{H0}
\end{equation}%
in the basis $\left\vert e_{i}\right\rangle $ and $\left\vert
s_{i}\right\rangle $, where each matrix element is the $N$-dimensional
matrix in the coordinate basis. In the notation $[G_{0}(k)]_{ij}^{\sigma
\sigma ^{\prime }}$, $\sigma (\sigma ^{\prime })=e,s$ denotes the different
\textquotedblleft spin\textquotedblright\ blocks and $i$, $j$ denote the
coordinate in the block. In the Markov limit $\omega d\ll 1$, $\mathcal{H}%
_{0}(\omega )$ is approximated by the frequency independent effective
Hamiltonian%
\begin{equation}
\mathcal{H}_{0}^{\mathrm{M}}=\left(
\begin{array}{cc}
(\Delta _{e}-i\Gamma _{f})\delta _{ij}-i\Gamma e^{ik_{0}\left\vert
x_{i}-x_{j}\right\vert } & \Omega \delta _{ij} \\
\Omega \delta _{ij} & \Delta _{s}\delta _{ij}%
\end{array}%
\right) ,  \label{H0M}
\end{equation}

In Fig. \ref{singlepolaron}, the exact transmission probability is compared
with that under the Markov approximation. As shown in Figs. \ref%
{singlepolaron}a and \ref{singlepolaron}b, for the small lattice spacing $%
d=10^{-4}$, the Markov approximation works very well, which leads to the
same result as that from Eq. (\ref{H0}). When the lattice spacing $d=10^{-2}$
is larger, Fig. \ref{singlepolaron}c shows the difference between the
results under the Markov approximation (\ref{H0M}) and the exact result (\ref%
{H0}). In the realistic case, the small length of the array justifies the
validity of the Markov approximation. Henceforth, we focus on the Markov
limit. In Figs. \ref{singlepolaron}b and \ref{singlepolaron}d, we have taken
into account the decay of the excited state $\left\vert e_{i}\right\rangle $
to the free space, i.e., $\Gamma _{f}\neq 0$. The single photon transmission
shows the EIT nature of atoms, where the total transmission appears at the
resonant frequency $k=0$ in the EIT window. The reason for the total
transmission is that the resonant photon transforms to the free-propagating
dark polariton excitation, which is only the superposition of states $%
r_{k}^{\dagger }\left\vert g_{i}\right\rangle $ and $\left\vert
s_{i}\right\rangle $. As a result, the free space decay $\Gamma _{f}$ of the
state $\left\vert e_{i}\right\rangle $ does not affect the total
transmission of the resonant dark polariton with $k=0$.

For the two incident photons with momenta $k_{1}$ and $k_{2}$ and the atoms
initially in the ground state, the initial boundary condition is $\gamma _{%
\mathrm{in}}=I$ and $\mathcal{F}_{\mathrm{in}}=\lim_{\{J_{k}\}\rightarrow
0}\delta ^{2}/\delta J_{k_{1},r}\delta J_{k_{2},r}$ in the asymptotic limit $%
t_{i}\rightarrow -\infty $. By the boundary condition of the final state $%
\gamma _{\mathrm{out}}=I$ and $\mathcal{F}_{\mathrm{out}}=\lim_{\{J_{k}\}%
\rightarrow 0}\delta ^{2}/\delta J_{p_{1},r}\delta J_{p_{2},r}$ in the
asymptotic limit and $t_{f}\rightarrow \infty $, Eqs. (\ref{At}) and (\ref%
{AJt}) lead to the two photon $S$-matrix%
\begin{eqnarray}
&&S_{p_{1}p_{2};k_{1}k_{2}}=T_{k_{1}}T_{k_{2}}(\delta _{p_{1}k_{1}}\delta
_{p_{2}k_{2}}+\delta _{p_{1}k_{2}}\delta _{p_{2}k_{1}})  \notag \\
&&+\frac{\Gamma ^{2}}{(2\pi )^{2}}%
\sum_{i_{1}i_{2},j_{1}j_{2}}e^{ik_{0}(x_{j_{1}}+x_{j_{2}}-x_{i_{1}}-x_{i_{2}})}
\\
&&\times G_{i_{1}i_{2};j_{1}j_{2}}^{ee;ee}(p_{1},p_{2};k_{1},k_{2})  \notag
\end{eqnarray}%
for two transmitted photons with momenta $p_{1}$ and $p_{2}$. Here, the
Green function%
\begin{eqnarray}
&&G_{i_{1}i_{2};j_{1}j_{2}}^{ee;ee}(p_{1},p_{2};k_{1},k_{2})  \notag \\
&=&\int dt_{1}^{\prime }dt_{2}^{\prime }dt_{1}dt_{2}e^{ip_{1}t_{1}^{\prime
}+ip_{2}t_{2}^{\prime }-ik_{1}t_{1}-ik_{2}t_{2}}  \notag \\
&&\times \left\langle \mathcal{T}e_{i_{1}}(t_{1}^{\prime
})e_{i_{2}}(t_{2}^{\prime })e_{j_{1}}^{\dagger }(t_{1})e_{j_{2}}^{\dagger
}(t_{2})\right\rangle _{c}
\end{eqnarray}%
is the Fourier transformation of the four-point connected Green function.

The Dyson expansion in terms of the two-body interaction $H_{\mathrm{HC}%
}+\sum_{ij}U_{ij}s_{i}^{\dagger }s_{j}^{\dagger }s_{j}s_{i}/2$ leads to%
\begin{eqnarray}
&&S_{p_{1}p_{2};k_{1}k_{2}}=S_{p_{1}p_{2};k_{1}k_{2}}^{(0)}-i\frac{\Gamma
^{2}}{2\pi }\delta _{p_{1}+p_{2},k_{1}+k_{2}}\times \\
&&\sum_{\substack{ iji^{\prime }j^{\prime }  \\ \sigma _{1}\sigma
_{1}^{\prime }\sigma _{2}\sigma _{2}^{\prime }}}[w^{\ast
}(p_{1},p_{2})]_{ij}^{\sigma _{1}\sigma _{1}^{\prime }}[T(E)]_{ij;i^{\prime
}j^{\prime }}^{\sigma _{1}\sigma _{1}^{\prime };\sigma _{2}\sigma
_{2}^{\prime }}[w(k_{1},k_{2})]_{i^{\prime }j^{\prime }}^{\sigma _{2}\sigma
_{2}^{\prime }}  \notag \\
&&+(p_{1}\leftrightarrow p_{2}),  \notag
\end{eqnarray}%
where we define the independent scattering part $%
S_{p_{1}p_{2};k_{1}k_{2}}^{(0)}=T_{k_{1}}T_{k_{2}}\delta _{p_{1}k_{1}}\delta
_{p_{2}k_{2}}$, and the function%
\begin{equation}
\lbrack w(k_{1},k_{2})]_{i^{\prime }j^{\prime }}^{\sigma \sigma ^{\prime
}}=\sum_{j_{1}j_{2}}e^{ik_{0}(x_{j_{1}}+x_{j_{2}})}[G_{0}(k_{1})]_{i^{\prime
}j_{1}}^{\sigma e}[G_{0}(k_{2})]_{j^{\prime }j_{2}}^{\sigma ^{\prime }e}.
\end{equation}

\begin{figure}[tbp]
\includegraphics[bb=17 230 584 763, width=8 cm, clip]{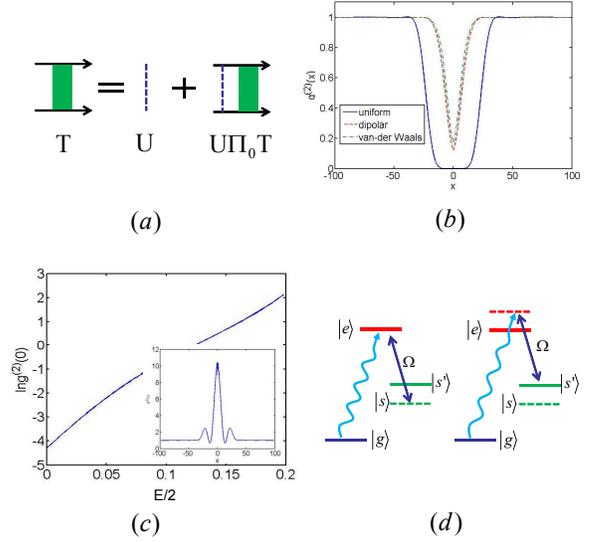}
\caption{(Color Online) Bunching and anti-bunching behaviors of transmitted
photons, where $\Gamma=1$, $\Gamma_{f}=1$, $\Delta_{e}=0$, $\Omega=1$, $%
k_{0}d=\protect\pi/2$, $U_{0}=10^{8}$, and the atom number is 20: (a) The
Feynman diagram for the $T $-matrix; (b) The second order correlation
functions $g^{(2)}(x)$ for $C=C_{3}=C_{6}=1$, $E=0$, and $k=0$; (c) The
second order correlation functions $\ln g^{(2)}(0)$ for the uniform case $%
C=0.46$, where the relative momentum $k=0$, and the inset shows $g^{(2)}(x)$
for the frequency $E/2=0.2$ of each photon; (d) The schematic for the
generation of the bunched and anti-bunched photons, where the incident
photons have different frequencies. $s'$ is the shifted energy level due to the Rydberg interaction.}
\label{g2tr}
\end{figure}

The $T$-matrix, depicted by the ladder diagram in Fig. \ref{g2tr}a,
satisfies the Lippmann-Schwinger equation%
\begin{eqnarray}
&&[T(E)]_{ij;i^{\prime }j^{\prime }}^{\sigma _{1}\sigma _{1}^{\prime
};\sigma _{2}\sigma _{2}^{\prime }}=U_{ij}^{\sigma _{1}\sigma _{1}^{\prime
}}\delta _{i^{\prime }i}\delta _{j^{\prime }j}\delta _{\sigma _{1}\sigma
_{2}}\delta _{\sigma _{1}^{\prime }\sigma _{2}^{\prime }} \\
&&+U_{ij}^{\sigma _{1}\sigma _{1}^{\prime }}\sum_{i_{1}j_{1};\mu _{1}\mu
_{1}^{\prime }}[\Pi (E)]_{ij;i_{1}j_{1}}^{\sigma _{1}\sigma _{1}^{\prime
};\mu _{1}\mu _{1}^{\prime }}[T(E)]_{i_{1}j_{1};i^{\prime }j^{\prime }}^{\mu
_{1}\mu _{1}^{\prime };\sigma _{2}\sigma _{2}^{\prime }},  \notag
\end{eqnarray}%
where the vacuum bubble is%
\begin{equation}
\lbrack \Pi (E)]_{ij;i^{\prime }j^{\prime }}^{\sigma \sigma ^{\prime };\mu
\mu ^{\prime }}=i\int \frac{d\omega }{2\pi }[G_{0}(\omega )]_{ii^{\prime
}}^{\sigma \mu }[G_{0}(E-\omega )]_{jj^{\prime }}^{\sigma ^{\prime }\mu
^{\prime }}.
\end{equation}%
In the matrix form, the Lippmann-Schwinger equation is formally solved as%
\begin{equation}
T(E)=\frac{1}{\mathbf{U}^{-1}-\Pi (E)},
\end{equation}%
where the vacuum bubble $\Pi _{0}(E)=(E-\mathcal{H}_{2})^{-1}$ is given by $%
\mathcal{H}_{2}=\mathcal{H}_{0}^{\mathrm{M}}\otimes I_{2N}+I_{2N}\otimes
\mathcal{H}_{0}^{\mathrm{M}}$, and the interaction matrix $\mathbf{U}$ has
the diagonal element $U_{ij}^{ee}=U_{ij}^{es}=U_{ij}^{se}=U_{0}\rightarrow
\infty $ and $U_{ij}^{ss}=U_{ij}$ in the basis $\{\left\vert
e_{i}e_{j}\right\rangle ,\left\vert e_{i}s_{j}\right\rangle ,\left\vert
s_{i}e_{j}\right\rangle ,\left\vert s_{i}s_{j}\right\rangle \}$.

The wavefunction%
\begin{eqnarray}
&&\psi (x_{c},x)=\int \frac{dp_{1}dp_{2}}{2\pi }%
S_{p_{1}p_{2};k_{1}k_{2}}e^{ip_{1}x_{1}+ip_{2}x_{2}}  \notag \\
&=&\frac{e^{iEx_{c}}}{2\pi }\{2T_{\frac{E}{2}+k}T_{\frac{E}{2}-k}\cos
(kx)-i\Gamma ^{2}\sum_{\substack{ iji^{\prime }j^{\prime }  \\ \sigma \sigma
^{\prime }\mu \mu ^{\prime }}}[F(x)]_{ij}^{\sigma \sigma ^{\prime }}  \notag
\\
&&\times \lbrack T(E)]_{ij;i^{\prime }j^{\prime }}^{\sigma \sigma ^{\prime
};\mu \mu ^{\prime }}[w(\frac{E}{2}+k,\frac{E}{2}-k)]_{i^{\prime }j^{\prime
}}^{\mu \mu ^{\prime }}\},  \label{wf2}
\end{eqnarray}%
of two transmitted photons is the Fourier transform of $%
S_{p_{1}p_{2};k_{1}k_{2}}$, where%
\begin{eqnarray}
&&[F(x)]_{ij}^{\sigma \sigma ^{\prime }}=-i\sum_{i_{1}i_{2}}\frac{%
e^{-ik_{0}x_{c,12}}}{E-\varepsilon _{l}-\varepsilon _{l^{\prime }}}%
\sum_{ll^{\prime }}\chi _{l}(i_{1}e)\chi _{l^{\prime }}(i_{2}e)  \notag \\
&&\tilde{\chi}_{l}^{\ast }(i\sigma )\tilde{\chi}_{l^{\prime }}^{\ast
}(j\sigma ^{\prime })[e^{i(\frac{E}{2}-\varepsilon _{l^{\prime }})x}\theta
(x)+e^{-i(\frac{E}{2}-\varepsilon _{l})x}\theta (-x)]  \notag \\
&&+(x\rightarrow -x).
\end{eqnarray}%
is determined by $x_{c,12}=x_{i_{1}}+x_{i_{2}}$ and the eigenstates $%
\left\vert \chi _{l}\right\rangle $ ($\left\vert \tilde{\chi}%
_{l}\right\rangle $) of $\mathcal{H}_{0}^{\mathrm{M}}$ ($\mathcal{H}_{0}^{%
\mathrm{M\dagger }}$) with the corresponding eigenenergies $\varepsilon _{l}$
($\varepsilon _{l}^{\ast }$). Here, $\left\langle i\sigma \left\vert \chi
_{l}\right\rangle \right. =\chi _{l}(i\sigma )$ and $\left\langle i\sigma
\left\vert \tilde{\chi}_{l}\right\rangle \right. =\tilde{\chi}_{l}(i\sigma )$%
.

By the wavefunction (\ref{wf2}), we show the normalized second order
correlation function $g^{(2)}(x)=\left\vert \pi \psi
(x_{c},x)/T_{E/2}^{2}\right\vert ^{2}$ of out-going photons for the two
incident photons with the same momentum $k_{1}=k_{2}=E/2$ in Figs. \ref{g2tr}%
b and \ref{g2tr}c. For the resonant case $E=0$, Fig. \ref{g2tr}b shows the
second order correlation functions for the uniform interaction $%
U_{ij}^{ss}=C $, the van-der Waals interaction $U_{ij}^{ss}=C_{6}/\left\vert
i-j\right\vert ^{6}$, and the dipolar interaction $U_{ij}^{ss}=C_{3}/\left%
\vert i-j\right\vert ^{3}$, which exhibit the anti-bunching behavior of
out-going photons.

It can be understood in the following way. As shown in the left panel of
Fig. \ref{g2tr}d, if two Rydberg excitations are close to each other, the
Rydberg state is shifted by the strong Rydberg interaction, such that the
classical light is off-resonance with respect to the transition between $%
\left\vert e_{i}\right\rangle $ and $\left\vert s_{i}\right\rangle $. As a
result, the photon with $k=0$ is resonant with the transition $\left\vert
g_{i}\right\rangle \rightarrow \left\vert e_{i}\right\rangle $ and
reflected. For two transmitted photons, to maintain that the frequency $k=0$
of the photon is in the EIT transmission window, they repulse each other and
show the anti-bunching behavior such that the Rydberg state is not shifted.

For the incident photons with frequency $E/2$ larger than some critical
value, the transmitted photons can also display bunching behavior, as shown
in Fig. \ref{g2tr}c. The mechanism of the generation of bunched photons is
illustrated in the right panel of Fig. \ref{g2tr}d. For two Rydberg
excitations close to each other, the Rydberg energy levels $\left\vert
s_{i}\right\rangle $ are shifted. The photons with finite momentum $E/2$
realize the two photon resonance with the shifted energy level $\left\vert
s_{i}\right\rangle $, and can be transmitted. In order to achieve the
transmission of photons by the shifted Rydberg levels $\left\vert
s_{i}\right\rangle $, the two Rydberg excitations prefer to stay next to
each other, which induces the bunching behavior of the transmitted photons.

As discussed in Sec. III, the second order correlation function $g^{(2)}(x)$
obtained by the scattering theory can characterize the quantum statistics of
photons emitted by the Rydberg-EIT atoms under the weak driving field. In
Ref. \cite{RydDarrick}, the result from the scattering theory and the
numerical solution of the master equation for the effective spin model under
the weak driving light are compared, where the two results agree with each
other perfectly.

\subsection{Propagations of single and two excitations}

In this section, we investigate how the single and two incident photons
transform to the excitations of the Rdyberg atoms, and the propagation of
excitations.

\begin{figure}[tbp]
\includegraphics[bb=10 502 576 759, width=8 cm, clip]{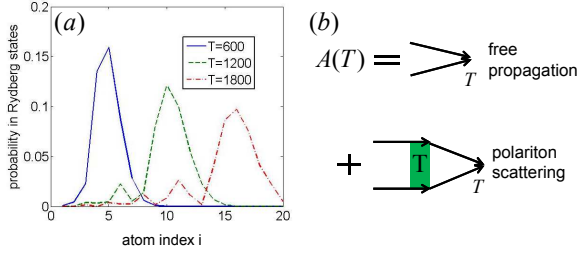}
\caption{(Color Online) (a) Probability of Rydberg states for the single
incident photon wavepacket, where $\Gamma=1$, $\Gamma_f=0$, $\Delta=0$, $%
\Omega=0.1$, $d=10^{-4}$, $\protect\gamma=0.01$, and the atom number is 20.
(b) The Feynman diagram for the two excitation propagation.}
\label{singlepro}
\end{figure}

For the single incident photon with the wavepacket $f(k)$ and the atoms
initially in the ground state, the boundary condition is $\gamma _{\mathrm{in%
}}=I$ and $\mathcal{F}_{\mathrm{in}}=\sum_{k}f_{(r,l)}(k)\lim_{\{J_{k}\}%
\rightarrow 0}\delta /\delta J_{k,(r,l)}$ at the instant $t_{i}=0$, where
the wavepackets of the right- and left- moving photons are%
\begin{equation}
f(k)=f_{r}(k)=\sqrt{\frac{\gamma }{\pi }}\frac{1}{k+i\gamma },
\end{equation}%
and%
\begin{equation}
f(k)=f_{l}(k)=\sqrt{\frac{\gamma }{\pi }}\frac{e^{-ikx_{N}}}{k-i\gamma },
\end{equation}
with the width $1/\gamma $. By the final boundary condition $\mathcal{F}_{%
\mathrm{out}}=1$ and $\gamma _{\mathrm{out}}=\mu _{i}=e_{i}$,$s_{i}$ at the
instant $t_{f}=T$, Eqs. (\ref{At}) and (\ref{AJt}) lead to the amplitude%
\begin{eqnarray}
\mathcal{A}_{i\mu }^{(\alpha )}(T) &=&i\sigma _{\alpha }\sqrt{2\gamma \Gamma
}\sum_{lj}e^{i\sigma _{\alpha }k_{0}x_{j}}\theta (T_{\alpha }-\sigma
_{\alpha }x_{j})\times \\
&&\frac{\chi _{l}(i\mu )\tilde{\chi}_{l}^{\ast }(je)}{\varepsilon
_{l}+i\gamma }[e^{-\gamma (T_{\alpha }-\sigma _{\alpha
}x_{j})}-e^{-i\varepsilon _{l}(T_{\alpha }-\sigma _{\alpha }x_{j})}],  \notag
\end{eqnarray}%
of the excitation $\mu =e,s$ at the position $i$, where $T_{\alpha
}=T-z_{\alpha }$ and $z_{r}=0,z_{l}=x_{N}$. For this situation, the dark
polariton forms, where the probability of the occupation in the excited
state $\left\vert e_{i}\right\rangle $ is quite small $\sim 10^{-4}$. In
Fig. \ref{singlepro}, we show the dark polariton propagation for the single
right moving incident photon. Tn order to show the slow propagation of the
dark polariton, we choose the small ratio $\Omega /\Gamma =0.1$.

For the wavepacket of two incident photons and the atoms initially in the
ground state, we consider both the co-propagation and counter- propagation
cases. For the co- propagation, the initial boundary condition at the
instant $t_{i}=0$ is $\gamma _{\mathrm{in}}=I$ and%
\begin{equation}
\mathcal{F}_{\mathrm{in}}=\sum_{k_{1}k_{2}}f_{r}(k_{1})f_{r}(k_{2})\lim_{%
\{J_{k}\}\rightarrow 0}\frac{\delta ^{2}}{\delta J_{k_{1},r}\delta
J_{k_{2},r}},
\end{equation}%
while for the counter- propagation, the initial boundary condition at the
instant $t_{i}=0$ is $\gamma _{\mathrm{in}}=I$ and%
\begin{equation}
\mathcal{F}_{\mathrm{in}}=\sum_{k_{1}k_{2}}f_{r}(k_{1})f_{l}(k_{2})\lim_{%
\{J_{k}\}\rightarrow 0}\frac{\delta ^{2}}{\delta J_{k_{1},r}\delta
J_{k_{2},l}}.
\end{equation}

By the boundary condition $\mathcal{F}_{\mathrm{out}}=1$ and $\gamma _{%
\mathrm{out}}=\mu _{1,i_{1}}\mu _{2,i_{2}}$ ($\mu _{1,2}=e,s$) at the
instant $t_{f}=T$, Eqs. (\ref{At}) and (\ref{AJt}) result in\ the amplitude%
\begin{eqnarray}
&&\mathcal{A}_{i_{1}i_{2};\mu _{1}\mu _{2}}^{(r,\alpha )}(T)=\mathcal{P}%
_{i_{1}\mu _{1};i_{2}\mu _{2}}\{\mathcal{A}_{i_{1}\mu _{1}}^{(r)}(T)\mathcal{%
A}_{i_{2}\mu _{2}}^{(\alpha )}(T)  \notag \\
&&+2\gamma \Gamma \sigma _{\alpha
}\sum_{j_{1}j_{2}}e^{ik_{0}x_{j_{1}}+i\sigma _{\alpha }k_{0}x_{j_{2}}}\theta
(T-x_{j_{1}})\times  \notag \\
&&\theta (T-z_{\alpha }-\sigma _{\alpha }x_{j_{2}})\int \frac{d\omega _{1}}{%
2\pi }\int \frac{d\omega _{2}}{2\pi }\times \\
&&\sum_{i^{\prime }j^{\prime },\sigma \sigma ^{\prime }}[\frac{1}{\omega
_{1}+\omega _{2}-\mathcal{H}_{2}}]_{i_{1}i_{2};ij}^{\mu _{1}\mu _{2};\sigma
\sigma ^{\prime }}\mathbf{U}_{ij}^{\sigma \sigma ^{\prime }}\times  \notag \\
&&[G(\omega _{1})]_{ij_{1}}^{\sigma e}[G(\omega _{2})]_{jj_{2}}^{\sigma
^{\prime }e}\frac{e^{-i\omega _{1}(T-x_{j_{1}})}}{\omega _{1}+i\gamma }\frac{%
e^{-i\omega _{2}(T-z_{\alpha }-\sigma _{\alpha }x_{j_{2}})}}{\omega
_{2}+i\gamma }\},  \notag
\end{eqnarray}%
of two excitations $\mu _{1}$ and $\mu _{2}$ at the positions $i_{1}$ and $%
i_{2}$ for the co-propagation $\alpha =r$ and the counter-propagation $%
\alpha =l$ case, where the operator $\mathcal{P}_{i_{1}\mu _{1};i_{2}\mu
_{2}}$ symmetrizes the wavefunction under the interchange $i_{1}\mu
_{1}\longleftrightarrow i_{2}\mu _{2}$.

\begin{figure}[tbp]
\includegraphics[bb=22 338 572 755, width=8 cm, clip]{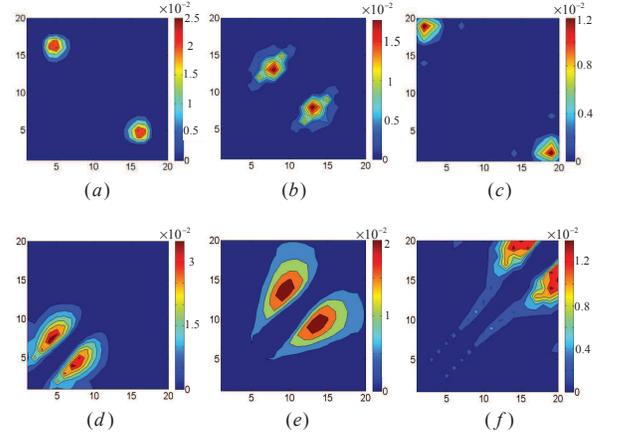}
\caption{(Color Online) Probabilities in the state $\left\vert
s\right\rangle $ for two counter- and co-propagating polaritons with the
dipolar case $C_{3}=1$, where $\Gamma=1$, $\Gamma_f=0$, $\Delta_{e}=0$, $%
d=10^{-4}$, and the atom number is 20. (a)-(c) show the probabilities at the
instants $T=600$, $1200$, and $1800$ for the counter-propagation case: $%
\Omega =0.1$ and $\protect\gamma =0.01$; (d)-(f) show the probabilities at
the instants $T=12$, $20$, and $28$ for the co-propagation case: $\Omega=1$
and $\protect\gamma =0.1$.}
\label{2p}
\end{figure}

In the first row of Fig. \ref{2p}, we show the propagation of two dark
polaritons for two counter-propagating incident photons. Here, we choose $%
\Omega /\Gamma =0.1$ to show the collision of two slow polaritons. When the
two excitations approach with each other, the states $\left\vert
s_{i}\right\rangle $ are shifted by the Rydberg interaction, which results
in the off-resonance with the transition from $\left\vert e_{i}\right\rangle
$ to $\left\vert s_{i}\right\rangle $ induced by the classical light. As a
result, the photon is resonant with the excited state $\left\vert
e_{i}\right\rangle $ and reflected. After the collision of two excitations,
they propagate away from each other, and the interaction gradually vanishes,
which results in the re-formation of two free propagating dark polaritons.

In the second row of Fig. \ref{2p}, the propagation of two dark polaritons
for the co-propagation incident photons is shown. When the first photon
transforms into the dark polariton in the atom array, it blocks the
transmission of the second photon. This blockade occurs over a
characteristic distance $r_{b}\sim (C_{3}\Gamma /\Omega ^{2})^{1/3}$. For an
incoming two-photon wavepacket whose size is larger than $r_{b}$, the
blockade manifests itself as a suppression of the probability of two photons
to overlap with each other as they propagate through the medium (see Fig. 10
d-f). During the propagation in the Rydberg-EIT atoms, the two polaritons
keep away from each other such that the energy level $\left\vert
s_{i}\right\rangle $ are not shifted, which results in the transmission of
two anti-bunched dark polaritons.

\section{Conclusion}

We summarize our results in this section. We developed the scattering theory
to investigate propagation of photons through an array of quantum emitters
using the path integral approach. The exact transition amplitude for
arbitrary initial and final states is obtained to describe the quantum
statistics of scattering photons and the dynamics of emitters in the
transient process. The exact result justifies the correctness of the Markov
approximation for the single emitter coupled to the waveguide photons with
linear dispersion. The exact and Markovian results coincide when the
bandwidth of the dynamics is sufficiently small compared to the distance
between emitters. Here, the generalized master equation for the few-photon
scattering process is obtained to describe the transient dynamics of
emitters. The generalized master equation establishes the relation between
two equivalent systems, i.e., few photon scattering by the emitters and the
emitters under weak driving light.

For the single emitter case, two paradigmatic examples, i.e., the two-level
emitter and the JC system, are used to show the correctness of our theory by
comparison with the well-known results. For an array of emitters, the
validity of the Markov approximation is examined in the system with
two-level emitters coupled to waveguide photons. Here, the dynamical
evolution of emitters show some non-Markovian effects, i.e., the retardation
effect.

The generation of entangled photons by the single emitter in front of the
mirror is also analyzed by the exact result from our theory. Finally, the
photon transmission in an array of EIT atoms coupled to the Rydberg level is
investigated by the scattering theory. We highlight the interplay between
the EIT phenomenon and the Rydberg interaction, and show how this results in
the bunching and anti-bunching behaviors of the scattering photons and the
dark polaritons propagating in the array.

The scattering theory also provides the way to explore some non-Markovian
effects of emitters coupled to the waveguide with nonlinear spectrum, where
the multi-photon bound states \cite{2b,2barray} may form. The general result
of the transition amplitude and the generalized master equation enable us to
study the photon transmission in more complicated quantum optics systems and
the dissipative many-body systems.

\acknowledgments This project has been supported by the EU project SIQS. DEC
acknowledges support from Fundacio Privada Cellex Barcelona, ERC Starting
Grant FOQAL, and the Ramon y Cajal program. Tao Shi thanks the useful
discussions with Y. Chang, A. G. Tudela, C. N. Benlloch, V. Paulisch, C. S.
Mu\~{n}oz, T. Caneva, and M. T. Manzoni.

\appendix

\section{Generalized master equation by path integral}

In this Appendix, we derive the generalized master equation (\ref{gr}) by
the path integral approach. To simplify the notation, we neglect the term $%
H_{\mathrm{\Gamma }_{f}}$ describing the independent decay of each emitter
to the free space, and the derivation including the free space mode follows
the same procedure. In the end, we show the result by considering the effect
of the free space decay.

Here, we need to introduce the emitter field explicitly. In quantum optics
systems, the emitter operators are usually the annihilation (creation)
operators of bosonic modes and the ladder operators. If one use the hardcore
boson to describe the emitters, all the operators are bosonic operators ($%
b_{l}$), where $l$ denotes the different modes of the emitters. In the
coherent basis, the reduced density operator $\rho _{s}(T)=Tr_{\mathrm{bath}%
}[e^{-iHT}\rho (0)e^{iHT}]$ becomes $\rho _{s}(T)=\mathcal{F}_{\mathrm{in}%
,-}^{\ast }\mathcal{F}_{\mathrm{in},+}e^{\sum_{k,\alpha }J_{k,\alpha ,%
\mathrm{-}}^{\ast }J_{k,\alpha ,\mathrm{+}}}\rho _{J}(T)$:%
\begin{eqnarray}
\rho _{J}(T) &=&\int d\mu (\beta _{l,+,\mathrm{out}},\beta _{l,+,\mathrm{out}%
}^{\ast })d\mu (\beta _{l,-,\mathrm{out}},\beta _{l,-,\mathrm{out}}^{\ast })
\notag \\
&&e^{-\sum_{k,\alpha }J_{k,\alpha ,\mathrm{-}}^{\ast }J_{k,\alpha ,\mathrm{+}%
}}\rho _{J}(\beta _{l,+,\mathrm{out}}^{\ast },\beta _{l,-,\mathrm{out}};T)
\notag \\
&&\times \left\vert \{\beta _{l,+,\mathrm{out}}\}\right\rangle \left\langle
\{\beta _{l,-,\mathrm{out}}\}\right\vert ,  \label{rJ}
\end{eqnarray}%
where $\alpha =r,l$ denotes the right- and left- moving photon modes,%
\begin{equation}
\mathcal{F}_{\mathrm{in,}\pm }=\lim_{\{J_{k\alpha }\}\rightarrow
0}\sum_{\{n_{k\alpha }\}}\psi _{\mathrm{in}}(\{n_{k\alpha }\})\prod_{k\alpha
}\frac{1}{\sqrt{n_{k\alpha }!}}\frac{\delta ^{n_{k\alpha }}}{\delta
J_{k,\alpha ,\pm }^{n_{k\alpha }}},
\end{equation}%
the measure for the unnormalized coherent state $\left\vert \{\beta
_{l}\}\right\rangle $ is $d\mu (\beta _{l},\beta _{l}^{\ast
})=\prod_{l}(e^{-\left\vert \beta _{l}\right\vert ^{2}}d^{2}\beta _{l}/\pi )$%
, and the element%
\begin{eqnarray*}
&&\rho _{J}(\beta _{l,+,\mathrm{out}}^{\ast },\beta _{l,-,\mathrm{out}%
};T)=\int d\mu (\beta _{l,\pm ,\mathrm{in}},\beta _{l,\pm ,\mathrm{in}%
}^{\ast })\times \\
&&Z_{J}(\beta _{l,+,\mathrm{out}}^{\ast },\beta _{l,-,\mathrm{out}};\beta
_{l,+,\mathrm{in}},\beta _{l,-,\mathrm{in}}^{\ast };T)\rho _{0}(\beta _{l,+,%
\mathrm{in}}^{\ast },\beta _{l,-,\mathrm{in}})
\end{eqnarray*}%
is given by the propagator%
\begin{eqnarray}
&&Z_{J}(\beta _{l,+,\mathrm{out}}^{\ast },\beta _{l,-,\mathrm{out}};\beta
_{l,+,\mathrm{in}},\beta _{l,-,\mathrm{in}}^{\ast };T)  \notag \\
&=&\int d\mu (J_{k,\alpha },J_{k,\alpha }^{\ast })  \label{ZJ} \\
&&\times \left\langle \{J_{k,\alpha }\}\right\vert \left\langle \{\beta
_{l,+,\mathrm{out}}\}\right\vert e^{-iHT}\left\vert \{\beta _{l,+,\mathrm{in}%
}\}\right\rangle \left\vert \{J_{k,\alpha ,+}\}\right\rangle  \notag \\
&&\times \left\langle \{J_{k,\alpha ,-}\}\right\vert \left\langle \{\beta
_{l,-,\mathrm{in}}\}\right\vert e^{iHT}\left\vert \{\beta _{l,-,\mathrm{out}%
}\}\right\rangle \left\vert \{J_{k,\alpha }\}\right\rangle  \notag
\end{eqnarray}%
in the closed time path. Here, we define $\rho _{0}(\beta _{l,+,\mathrm{in}%
}^{\ast },\beta _{l,-,\mathrm{in}})=\left\langle \{\beta _{l,+,\mathrm{in}%
}\}\right\vert \rho _{\mathrm{sys}}(0)\left\vert \{\beta _{l,-,\mathrm{in}%
}\}\right\rangle $.

The propagator (\ref{ZJ}) is obtained by the saddle point method shown in
Sec. IIIC, which is%
\begin{eqnarray}
&&Z_{J}(\beta _{l,+,\mathrm{out}}^{\ast },\beta _{l,-,\mathrm{out}};\beta
_{l,+,\mathrm{in}},\beta _{l,-,\mathrm{in}}^{\ast };T)  \notag \\
&=&e^{iS_{b}}\times \int d\mu (J_{k,\alpha },J_{k,\alpha }^{\ast })  \notag
\\
&&\exp \{\sum_{k,\alpha =r,l}[J_{k,\alpha }^{\ast }J_{k,\alpha ,\mathrm{+}%
}e^{-i\sigma _{\alpha }kT}+J_{k,\alpha ,\mathrm{-}}^{\ast }J_{k,\alpha
}e^{i\sigma _{\alpha }kT}]\}  \notag \\
&&\times \int D[\mathrm{system}]e^{iS_{\mathrm{eff,+}}-iS_{\mathrm{eff,-}%
}^{\ast }+iS_{J,\mathrm{+}}-iS_{J,\mathrm{-}}^{\ast }}.
\end{eqnarray}%
Here, $iS_{b}=\sum_{l}[\beta _{l,+,\mathrm{out}}^{\ast }\beta _{l}(T)+\beta
_{l}^{\ast }(T)\beta _{l,-,\mathrm{out}}]$ is the boundary term for the
emitters, and $S_{\mathrm{eff,\pm }}$ are the effective actions in the
forward and backward time-evolution paths, which are given by substituting
the fields $\beta _{l,\pm }$ for the emitter fields in the effective action $%
S_{\mathrm{eff}}$. The external source term is%
\begin{eqnarray}
S_{J,\mathrm{\pm }} &=&-\int_{0}^{T}dt\sum_{k,\alpha }[J_{k,\alpha }^{\ast
}O_{k,\sigma _{\alpha },\mathrm{\pm }}(t)e^{-i\sigma _{\alpha }k(T-t)}
\notag \\
&&+J_{k,\alpha ,\mathrm{\pm }}O_{k,\sigma _{\alpha },\mathrm{\pm }}^{\ast
}(t)e^{-i\sigma _{\alpha }kt}].
\end{eqnarray}%
Finally, the Gaussian integral over $J_{k,\alpha },J_{k,\alpha }^{\ast }$
leads to the propagator%
\begin{eqnarray}
&&Z_{J}(\beta _{l,+,\mathrm{out}}^{\ast },\beta _{l,-,\mathrm{out}};\beta
_{l,+,\mathrm{in}},\beta _{l,-,\mathrm{in}}^{\ast };T)  \notag \\
&=&e^{\sum_{k,\alpha }J_{k,\alpha ,\mathrm{-}}^{\ast }J_{k,\alpha ,\mathrm{+}%
}}\int D[\mathrm{system}]e^{i(S_{\mathrm{d,+}}-S_{\mathrm{d,-}}+S_{\mathrm{%
jump}})}  \notag \\
&&\exp \{\sum_{l}[\beta _{l,+,\mathrm{out}}^{\ast }\beta _{l}(T)+\beta
_{l}^{\ast }(T)\beta _{l,-,\mathrm{out}}]\},  \label{zJ}
\end{eqnarray}%
where%
\begin{eqnarray}
S_{\mathrm{d,+}} &=&S_{\mathrm{eff,+}}-\int_{0}^{T}dt\sum_{k,\alpha
}[J_{k,\alpha ,\mathrm{-}}^{\ast }O_{k,\sigma _{\alpha },\mathrm{+}%
}(t)e^{i\sigma _{\alpha }kt}  \notag \\
&&+J_{k,\alpha ,\mathrm{+}}O_{k,\sigma _{\alpha },\mathrm{+}}^{\ast
}(t)e^{-i\sigma _{\alpha }kt}],
\end{eqnarray}%
\begin{eqnarray}
S_{\mathrm{d,-}} &=&S_{\mathrm{eff,-}}^{\ast }-\int_{0}^{T}dt\sum_{k,\alpha
}[J_{k,\alpha ,\mathrm{-}}^{\ast }O_{k,\sigma _{\alpha },\mathrm{-}%
}(t)e^{i\sigma _{\alpha }kt}  \notag \\
&&+J_{k,\alpha ,\mathrm{+}}O_{k,\sigma _{\alpha },\mathrm{-}}^{\ast
}(t)e^{-i\sigma _{\alpha }kt}],
\end{eqnarray}%
and the jump term%
\begin{equation}
S_{\mathrm{jump}}=-i\int dtdt^{\prime }\sum_{k,\alpha }e^{i\sigma _{\alpha
}k(t-t^{\prime })}O_{k,\sigma _{\alpha },\mathrm{+}}(t)O_{k,\sigma _{\alpha
},\mathrm{-}}^{\ast }(t^{\prime }).
\end{equation}

Under the Markov approximation, the actions read%
\begin{eqnarray}
S_{\mathrm{eff,\pm }} &=&S_{\mathrm{sys,\pm }}+i\int_{t_{i}}^{t_{f}}dt%
\sum_{ij}\sqrt{\Gamma _{i}\Gamma _{j}}  \notag \\
&&\times O_{i\mathrm{,\pm }}^{\ast }(t)O_{j\mathrm{,\pm }}(t)e^{ik_{0}\left%
\vert x_{i}-x_{j}\right\vert },
\end{eqnarray}%
and%
\begin{eqnarray}
S_{\mathrm{jump}} &=&-2i\int_{0}^{T}dt\sum_{ij}\sqrt{\Gamma _{i}\Gamma _{j}}%
\cos [k_{0}(x_{i}-x_{j})]  \notag \\
&&\times O_{i,\mathrm{+}}(t)O_{j,\mathrm{-}}^{\ast }(t).
\end{eqnarray}%
By taking the time derivative of Eq. (\ref{rJ}) and using Eq. (\ref{zJ}), we
obtain the motion equation%
\begin{equation}
\partial _{T}\rho _{J}(T)=-i[\tilde{H}_{\mathrm{sys}}(T),\rho _{J}(T)]+%
\mathcal{L}\rho _{J}(T)  \label{GME}
\end{equation}%
for the generating density matrix $\rho _{J}(T)$ with the initial condition $%
\rho _{J}(0)=\rho _{\mathrm{sys}}(0)$, where the Lindblad term is%
\begin{eqnarray}
&&\mathcal{L}\rho _{J}(T)=2\sum_{ij}\sqrt{\Gamma _{i}\Gamma _{j}}O_{i}\rho
_{J}(T)O_{j}^{\dagger }\cos [k_{0}(x_{i}-x_{j})]  \notag \\
&&-\sum_{ij}\sqrt{\Gamma _{i}\Gamma _{j}}\cos
[k_{0}(x_{i}-x_{j})]\{O_{i}^{\dagger }O_{j},\rho _{J}(T)\}.
\end{eqnarray}

Finally, by adding the Lindblad term describing the free space decay, we
reproduce the results (\ref{gr}), (\ref{hd}), and (\ref{ld}).

\end{document}